\documentclass[journal,twoside,web]{ieeecolor}
\usepackage{tmi}
\usepackage{cite}
\usepackage{amsmath,amssymb,amsfonts}
\usepackage{algorithmicx}
\usepackage{graphicx}
\usepackage{caption}
\usepackage{stfloats}
\usepackage{amsmath}
\usepackage{verbatim}
\usepackage{amsmath}
\usepackage{algorithm}
\usepackage{algpseudocode}

\def\BibTeX{{\rm B\kern-.05em{\sc i\kern-.025em b}\kern-.08em
    T\kern-.1667em\lower.7ex\hbox{E}\kern-.125emX}}
\begin{document}
\title{Rapid Bone Scintigraphy Enhancement via Semantic Prior Distillation from Segment Anything Model}

\author{Pengchen Liang,
Leijun Shi,
Huiping Yao,
Bin Pu,
Jianguo Chen, ~\IEEEmembership{Member, IEEE},
Lei Zhao,
Haishan Huang,
Zhuangzhuang Chen,
Zhaozhao Xu,
Lite Xu,
Qing Chang,
and Yiwei Li
\thanks{This work is partially funded by the Shanghai University of Engineering Science Medical-Engineering Interdisciplinary Project (2023LXY-RUIJIN01Z), and the Shanghai Children's Hospital Clinical Research Cultivation Project (Grant No.2022YLYM13).}
\thanks{Corresponding authors: Yiwei Li (liyiwei@shchildren.com.cn), Bin Pu (eebinpu@ust.hk) and Qing Chang (robie0510@hotmail.com)}
\thanks{Pengchen Liang is with the Department of Nuclear Medicine, Shanghai Children's Hospital, School of Medicine, Shanghai Jiao Tong University, Shanghai, 200062, China, and also with the School of Microelectronics, Shanghai University, Shanghai, 201800, China;}
\thanks{Leijun Shi, Lite Xu, and Yiwei Li are with the Department of Nuclear Medicine, Shanghai Children's Hospital, School of Medicine, Shanghai Jiao Tong University, Shanghai, 200062, China;}
\thanks{Huiping Yao is with Department of Ophthalmology, Ruijin Hospital, Shanghai Jiao Tong University School of Medicine, Shanghai, 201800, China}
\thanks{Bin Pu and Zhuangzhuang Chen are with the Department of Electronic and Computer Engineering, The Hong Kong University of Science and Technology, 999077, Hong Kong, China;}
\thanks{Jianguo Chen and Haishan Huang are with the School of Software Engineering, Sun Yat-sen University, Zhuhai, Guangdong Province, 519000, China;}
\thanks{Lei Zhao is with the College of Computer Science and Electronic Engineering, Hunan University, 410082, Changsha, China;}
\thanks{Zhaozhao Xu is with the School of Computer Science and Technology, Henan Polytechnic University, Jiaozuo, Henan 454000, China;}
\thanks{Qing Chang is with the Department Shanghai Key Laboratory of Gastric Neoplasms, Department of Surgery, Shanghai Institute of Digestive Surgery, Ruijin Hospital, Shanghai Jiao Tong University School of Medicine, Shanghai, 201800, China}
\thanks{Pengchen Liang, Leijun Shi, Huiping Yao, and Jianguo Chen contributed equally to this work.}}

\maketitle
\begin{abstract}
Rapid bone scintigraphy is crucial for diagnosing skeletal disorders and detecting tumor metastases in children, as it shortens scan duration and reduces discomfort. However, accelerated acquisition often degrades image quality, impairing the visibility of fine anatomical details and potentially compromising diagnosis. To overcome this limitation, we introduce the first application of SAM-based semantic priors for medical image restoration, utilizing the Segment Anything Model (SAM) to enhance pediatric rapid bone scintigraphy. Our approach employs two cascaded networks, $f^{IR1}$ and $f^{IR2}$, supported by three specialized modules: a Semantic Prior Integration (SPI) module, a Semantic Knowledge Distillation (SKD) module, and a Semantic Consistency Module (SCM). The SPI and SKD modules inject domain-specific semantic cues from a fine-tuned SAM, while the SCM preserves coherent semantic feature representations across both cascaded stages. Moreover, we present RBS, a novel Rapid Bone Scintigraphy dataset comprising paired standard (20 cm/min) and rapid (40 cm/min) scans from 137 pediatric patients aged 0.5–16 years, making it the first dataset tailored for pediatric rapid bone scintigraphy restoration. Extensive experiments on both a public endoscopic dataset and our RBS dataset demonstrate that our method consistently surpasses existing techniques in PSNR, SSIM, FID, and LPIPS metrics.

\end{abstract}


\section{Introduction}

Rapid bone scintigraphy using single-photon emission computed tomography (SPECT) provides significant advantages over traditional scanning methods, primarily by reducing scan time and minimizing patient discomfort. 
These benefits are particularly important for young patients who may struggle with prolonged immobility during standard scans \cite{murata2024verification,isherwood2023sub}.
However, the acceleration of scanning speed in rapid bone scintigraphy often results in degraded image quality. 
This degradation poses a significant challenge to diagnostic accuracy, potentially compromising the detection and assessment of skeletal abnormalities \cite{pan2024fast, moncayo2021can, alqahtani2023optimising}.
The tension between scan speed and image quality presents a critical area for improvement in pediatric bone scintigraphy. This challenge necessitates innovative approaches to restore the quality of rapid scan images without sacrificing the benefits of reduced bone scan time \cite{ito2022adapting,pan2022ultra,bahloul2024ultra}.

Deep learning-based methods, particularly Convolutional Neural Networks (CNNs), have shown remarkable success in image restoration tasks \cite{dong2015image, zhang2017beyond}, including bone scintigraphy image denoising \cite{ito2022adapting, chen2017low}. These approaches have demonstrated superior performance in noise reduction and detail preservation compared to conventional techniques \cite{qi2023deep, murata2024verification}.
Recent advancements emphasize the importance of incorporating semantic information into image restoration processes. Semantic segmentation, which involves partitioning an image into meaningful regions, has been shown to improve the accuracy and quality of restored images \cite{jiang2023restore,xiao2023dive}. 
In medical imaging, this approach helps preserve critical anatomical structures during restoration \cite{huang2022segmentation,yin2022segmentation}.

Recently, the Segment Anything Model (SAM) has demonstrated remarkable capabilities in various computer vision tasks \cite{kirillov2023segment}. 
SAM encapsulates rich semantic knowledge, offering potential for advanced image understanding and processing \cite{wu2023medical, hu2023efficiently,li2024adapting,cheng2024unleashing,zhang2024sam}.
Despite its promise, SAM has not yet been used to enhance the resolution of medical images, particularly in rapid bone scintigraphy.
Medical images often contain higher levels of noise and artifacts due to the nature of imaging processes \cite{sagheer2020review,goyal2018noise,florkow2022magnetic}, such as the high-speed scanning protocols used in rapid bone scintigraphy. 
These images also require the preservation of critical anatomical structures to ensure accurate diagnosis, making the task of image restoration more complex compared to general image restoration \cite{katsaggelos2012digital,yang2024all}.
In contrast to typical images, which may focus on aesthetic or perceptual qualities, medical images, such as bone scintigraphy scans, emphasize the accurate representation of anatomical features and pathology. 
The presence of low contrast and noise in rapid bone scintigraphy images complicates the restoration process. These images need to maintain fine details, such as bone structures, while removing noise and artifacts, which requires an advanced semantic understanding of the image content.

In our study, we propose a novel method that leverages SAM's semantic priors to enhance the quality of pediatric rapid bone scintigraphy images without compromising scan speed. 
Our approach integrates SAM's semantic knowledge into a specialized image restoration framework, specifically tailored to address the unique challenges inherent in rapid bone scintigraphy. Unlike existing SAM-based image restoration techniques designed for general-purpose imagery, our method involves fine-tuning SAM on a limited set of medically annotated data to generate precise semantic priors. This adaptation ensures that the semantic information guiding the restoration process is highly relevant and accurate for medical diagnostics.

Previous research in medical image restoration has been constrained by the scarcity of suitable datasets and the complexity of maintaining diagnostic integrity during image enhancement. To overcome these barriers, we introduce the first comprehensive dataset specifically designed for rapid bone scintigraphy image restoration in pediatric patients. This dataset comprises a diverse range of cases and scan parameters, providing a robust foundation for developing and evaluating innovative restoration techniques in this critical medical imaging domain.
The main contributions of this paper are as follows:

\begin{itemize}
\item We propose a SAM semantic-guided approach for deep learning-based image restoration, integrating fine-tuned SAM semantic masks to effectively mitigate noise and artifacts in rapid bone scintigraphy scans.
\item We develop a dual-stage distillation network comprising three key modules: Semantic Prior Integration, Semantic Knowledge Distillation, and a Semantic Consistency Module. This architecture enhances initial restoration quality, refines model performance, and ensures consistent semantic feature representation throughout the network.
\item  We release the Rapid Bone Scintigraphy (RBS) dataset, the first comprehensive resource specifically designed for pediatric rapid bone scintigraphy image restoration. 
RBS includes scans from 137 patients aged 0.5 to 16 years, addressing a critical gap in the field and facilitating advanced research in pediatric nuclear medicine imaging.
\item We demonstrate our method's versatility by successfully applying it to low-quality endoscopic surgical images, illustrating its potential for broader applications in medical image restoration beyond bone scintigraphy.
\end{itemize}

\section{Related work}

\subsection{Deep Learning in Medical Image Restoration}

Deep learning techniques have revolutionized the field of medical image restoration, particularly in addressing challenges associated with low-dose imaging modalities \cite{deng2024unsupervised,li2022annotation,geng2021content}. Convolutional Neural Networks (CNNs) have been at the forefront of this revolution, demonstrating remarkable capabilities in image super-resolution and denoising tasks \cite{dong2015image,chen2017low,zhang2017beyond,wu2022arbitrary}. 
Recently, transformer architectures, which have shown remarkable success in natural language processing, have been adapted for image recognition tasks \cite{dosovitskiy2020image,zhang2021transct}. 
In the specific context of nuclear medicine imaging, deep learning methods have been successfully applied to SPECT bone imaging \cite{pan2022ultra,qi2023deep}, demonstrating the feasibility of enhancing ultra-fast SPECT/CT bone scans and achieving comparable image quality and diagnostic value to standard acquisitions with significantly reduced scan times.

\subsection{Semantic Priors for Image Restoration}

Recent works have demonstrated the effectiveness of incorporating semantic priors into image restoration tasks \cite{jiang2023restore,zhang2024distilling}. One study introduced a human-aware deblurring model that disentangles motion blur between foreground humans and background, utilizing a human-aware attention mechanism to focus on specific domains \cite{shen2019human}. 
Another study explored the potential of SAM in image restoration, proposing a lightweight SAM prior tuning (SPT) unit that can be integrated into existing restoration networks \cite{xiao2023dive}.
In the medical imaging domain, the incorporation of segmentation information as domain knowledge for low-dose CT denoising has shown significant benefits \cite{huang2022segmentation, yin2022segmentation}.

\subsection{Foundation Models for Segmentation Tasks}

Foundation models have revolutionized the field of image segmentation, with SAM emerging as a groundbreaking development \cite{kirillov2023segment}. 
The success of SAM has inspired numerous adaptations and applications in specialized domains \cite{miao2024cross,wang2024sam,zhang2023customized}, particularly in medical imaging \cite{li2024carotid,gu2024lesam,qiu2023large,jiang2024glanceseg}. 
A notable example is the introduction of MedSAM, a foundation model for universal medical image segmentation, developed on a large-scale medical image dataset covering 10 imaging modalities and over 30 cancer types \cite{ma2024segment}.
Furthermore, various strategies have been proposed to enhance the performance of SAM in medical applications. One approach involves freezing SAM's encoder and fine-tuning a lightweight task-specific prediction head, which has demonstrated improved performance even with limited labeled data \cite{hu2023efficiently}. 
These developments underscore the growing importance and versatility of foundation models like SAM in advancing image segmentation across various domains, particularly in medical imaging.

\begin{figure*}[!ht]
\setlength{\abovecaptionskip}{0pt}
\setlength{\belowcaptionskip}{0pt}
\centering
\includegraphics[scale=0.62]{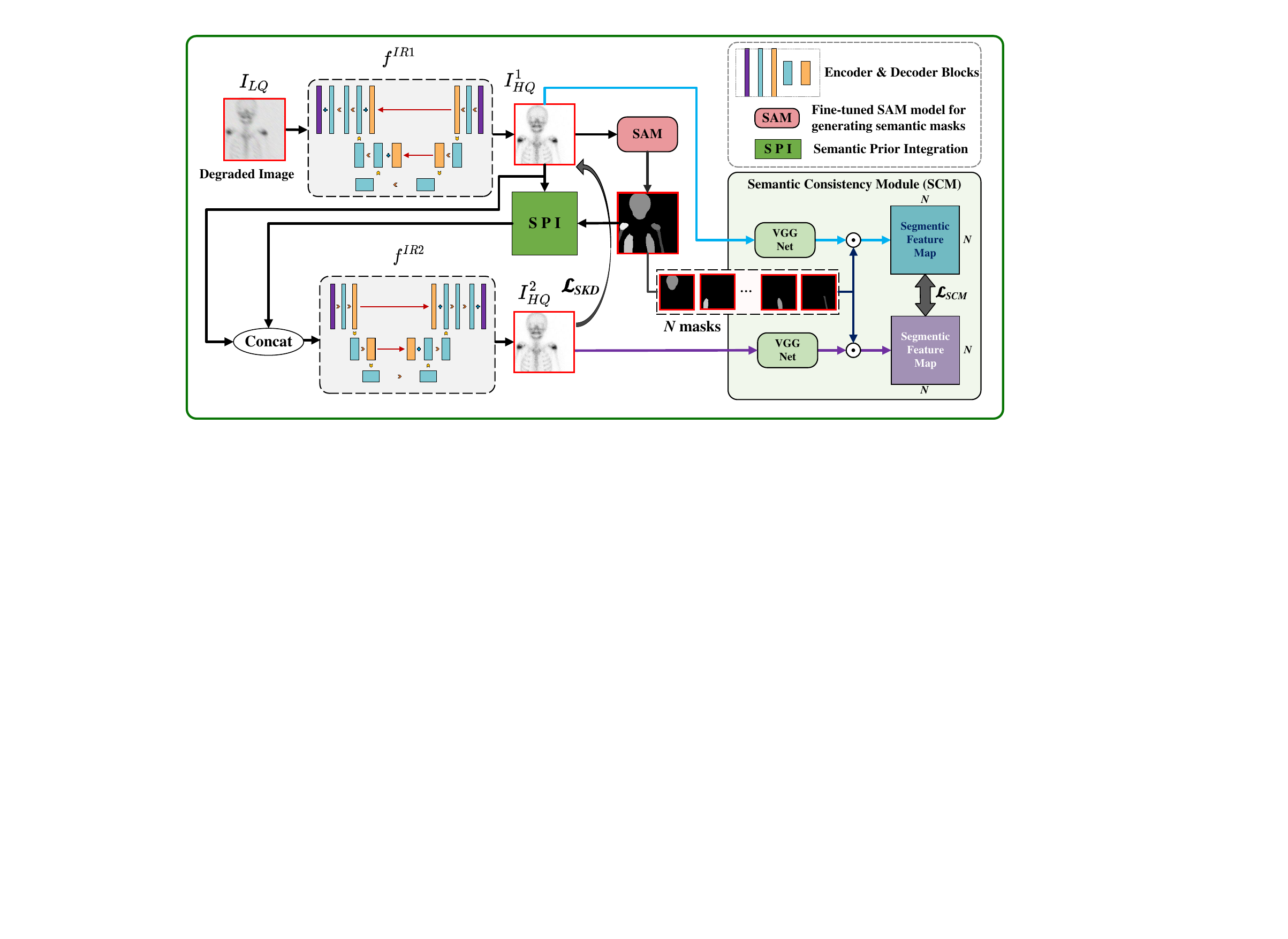}
\caption{Architecture of the proposed framework for enhancing rapid bone scintigraphy images. The framework consists of three main modules: (1) Semantic Prior Integration (SPI) module, which incorporates semantic information from SAM into the $f^{IR2}$ ; (2) Semantic Knowledge Distillation (SKD) module, which transfers knowledge from the enhanced $f^{IR2}$ to the original $f^{IR1}$; and (3) the Semantic Consistency Module (SCM), which ensures consistent semantic feature representation across the networks. This synergistic combination of deep learning and semantic priors significantly improves the image quality of rapid bone scans.}
\vspace{-.1in}
\label{fig1}
\end{figure*}

\section{Method}

\subsection{Overview}

We present a novel framework, SPD-VFM, for enhancing rapid bone scintigraphy images in pediatric patients by pioneering the integration of SAM's semantic priors with deep learning techniques in medical image restoration (Figure \ref{fig1}). Our method comprises three key synergistic components: Semantic Prior Integration (SPI), Semantic Knowledge Distillation (SKD), and Semantic Consistency Module (SCM). Additionally, the framework employs a two-stage distillation network involving two cascaded image restoration models, $f^{IR1}$ and $f^{IR2}$.
 
\textbf{Training Phase:} 
During training, the degraded low-quality (LQ) bone scintigraphy image, $I_{LQ}$, is first processed by the initial image restoration model, $f^{IR1}$, to produce an intermediate high-quality (HQ) image, $I_{HQ}^1$. Concurrently, a fine-tuned Segment Anything Model (SAM) generates semantic masks, $M$, from the ground truth image, $I_{HQ}$. These masks encapsulate crucial anatomical structures necessary for accurate restoration. To ensure stability and effectiveness in the two-stage distillation process, the second image restoration model, $f^{IR2}$, is \textbf{pre-trained independently} using the semantic information integrated by the SPI module with SAM-generated semantic masks, on both the low-quality image $I_{LQ}$ and the ground truth image $I_{HQ}$.

After pre-training, $f^{IR2}$ is incorporated into the two-stage network with the SPI module. Specifically, within the SPI module, a Mask Average Pooling (MAP) unit integrates the semantic masks $M$ into the features extracted by $f^{IR2}$ from the intermediate high-quality image $I_{HQ}^1$, resulting in a further refined high-quality output, denoted as $I_{HQ}^2$. The Semantic Knowledge Distillation (SKD) module facilitates the transfer of semantic knowledge from the pre-trained $f^{IR2}$ (Teacher) to the primary restoration model $f^{IR1}$ (Student) by minimizing the SKD loss, $\mathcal{L}_{SKD}$. This process enables $f^{IR1}$ to implicitly learn semantic features, enhancing its restoration capabilities without introducing additional computational complexity during inference.

Complementing these components, the Semantic Consistency Module (SCM) ensures that the semantic relationships between $I_{HQ}^1$ and $I_{HQ}^2$ remain consistent. SCM leverages a pre-trained VGG network to extract and align semantic feature representations, maintaining the structural integrity of anatomical features throughout the restoration process.

\textbf{Inference Phase:} 
Crucially, only the primary image restoration model, $f^{IR1}$, is utilized during inference. The second model, $f^{IR2}$, along with the SPI and SKD modules, remains inactive, ensuring that the computational overhead introduced by the two-stage training process does not affect deployment efficiency. This design choice allows the framework to deliver high-quality restorations comparable to the two-stage training approach without incurring additional computational costs during deployment.


\subsection{Problem Formulation}

Rapid bone scintigraphy scans, while efficient, often result in degraded image quality due to accelerated acquisition times. These images typically suffer from increased noise, motion artifacts, and loss of fine details, which can impede accurate diagnosis. Our primary objective is to enhance the original Image Restoration (IR) model, denoted as $f^{IR1}$, by leveraging semantic information through a secondary model $f^{IR2}$.

Let $\Omega \subset \mathbb{R}^2$ be the image domain. We define the degraded low-quality (LQ) bone scintigraphy image as $I_{LQ}: \Omega \to \mathbb{R}$, and our goal is to restore a high-quality (HQ) image $I_{HQ}: \Omega \to \mathbb{R}$. The degradation process can be modeled as:

\begin{equation}
I_{LQ} = \mathcal{D}(I_{HQ}) + \eta,
\end{equation}
where $\mathcal{D}: \mathbb{R}^{\Omega} \to \mathbb{R}^{\Omega}$ is a degradation operator that encompasses various factors such as blur, downsampling, and other artifacts, and $\eta$ represents additive noise, typically assumed to follow a Gaussian distribution $\eta \sim \mathcal{N}(0, \sigma^2)$.

Our restoration process involves two stages: initial restoration using $f^{IR1}$, followed by enhancement through $f^{IR2}$ that integrates semantic priors. However, our primary focus is on optimizing $f^{IR1}$. We can formulate this as an optimization problem:

\begin{equation}
\resizebox{0.96\hsize}{!}{$
\begin{aligned}
f^{IR1*} = \min_{f^{IR1}} \mathbb{E}_{I_{LQ}, I_{HQ}} \left[ \mathcal{L}(f^{IR2}(f^{IR1}(I_{LQ})), I_{HQ}) + \lambda \mathcal{R}(f^{IR1}) \right],
\end{aligned}
$}
\end{equation}
where $\mathcal{L}$ is a loss function measuring the discrepancy between the restored and ground truth images, $\mathcal{R}$ is a regularization term to prevent overfitting of $f^{IR1}$, and $\lambda$ is a regularization parameter.

To incorporate semantic information, we introduce a semantic prior $S: \Omega \to \{1, ..., K\}$, where $K$ is the number of semantic classes, obtained using a fine-tuned SAM. We then refine our optimization problem:

\begin{equation}
\begin{split}
f^{IR1*} = \min_{f^{IR1}} \mathbb{E}_{I_{LQ}, I_{HQ}, S} [ \mathcal{L}(f^{IR2}(f^{IR1}(I_{LQ}), S), I_{HQ}) + \\
\lambda_1 \mathcal{R}(f^{IR1}) + \lambda_2 \mathcal{C}(f^{IR1}(I_{LQ}), f^{IR2}(f^{IR1}(I_{LQ}), S)) ],
\end{split}
\end{equation}
where $\mathcal{C}$ is a consistency term ensuring that the output of $f^{IR1}$ aligns with the semantically enhanced output of $f^{IR2}$, and $\lambda_1, \lambda_2$ are weighting parameters.

To formalize the SPI in $f^{IR2}$, we define a Mask Average Pooling (MAP) operation $\mathcal{M}: \mathbb{R}^{\Omega} \times \{1, ..., K\}^{\Omega} \to \mathbb{R}^{\Omega}$ as:

\begin{equation}
\mathcal{M}(F, S)_p = \frac{1}{|\{q \in \Omega : S(q) = S(p)\}|} \sum_{q: S(q) = S(p)} F_q,
\end{equation}
where $F$ represents the feature map extracted by the encoder of $f^{IR2}$, and $p, q$ are pixel locations.

The restoration process can be expressed as:

\begin{equation}
I_{HQ}^1 = f^{IR1}(I_{LQ}),
\end{equation}
\begin{equation}
I_{HQ}^2 = f^{IR2}(\text{Concat}[\mathcal{M}(F_{enc}(I_{HQ}^1), S), I_{HQ}^1]),
\end{equation}
where $F_{enc}$ is the encoder of $f^{IR2}$, and Concat denotes the concatenation operation. 

The key insight is that while $f^{IR2}$ produces $I_{HQ}^2$, our primary goal is to optimize $f^{IR1}$ such that $I_{HQ}^1$ approaches the quality of $I_{HQ}^2$. This is achieved through the consistency term $\mathcal{C}$ in our optimization objective, which encourages $f^{IR1}$ to learn from the semantically enhanced results of $f^{IR2}$.

Our approach focuses on optimizing of $f^{IR1}$ while leveraging the semantic information integrated in $f^{IR2}$. It allows for a comprehensive analysis of the restoration process, taking into account the inherent noise, the semantic information, and the knowledge transfer from $f^{IR2}$ to $f^{IR1}$.

\subsection{Semantic Prior Integration}

\begin{figure}[!ht]
\setlength{\abovecaptionskip}{0pt}
\setlength{\belowcaptionskip}{0pt}
\centering
\includegraphics[scale=0.417]{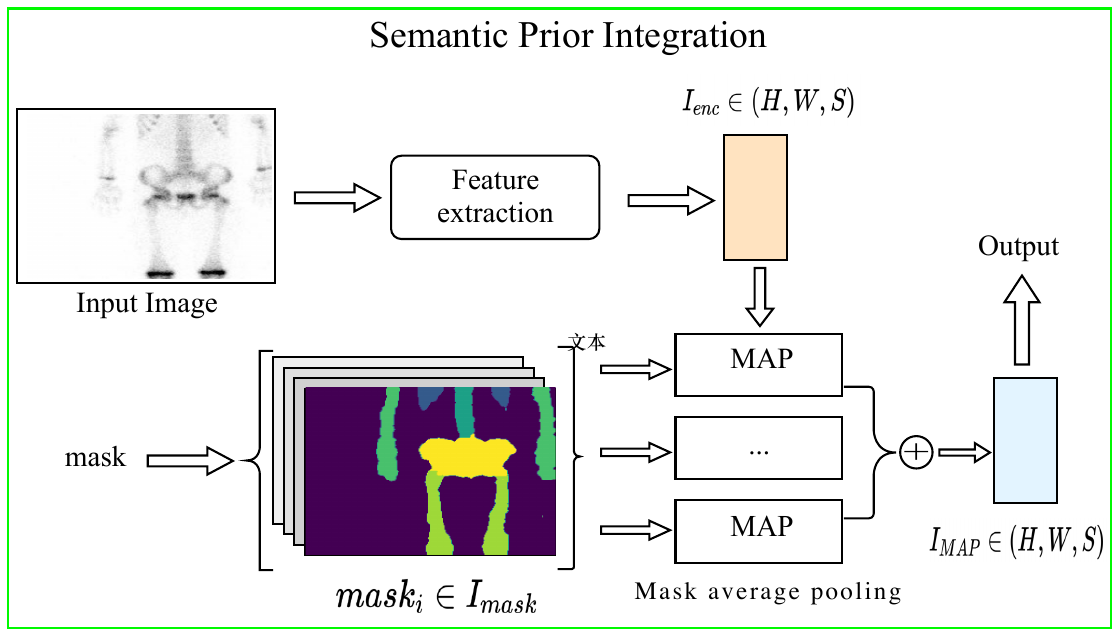}
\caption{Architecture of the Semantic Prior Integration (SPI) component. The input features ($I_{enc}$) are processed through multiple Mask Average Pooling (MAP) units guided by SAM-generated anatomical masks ($mask_i$). The outputs are fused by element-wise addition to produce semantically-enhanced features ($I_{MAP}$).}
\vspace{-.1in}
\label{fig:spi_architecture}
\end{figure}

The SPI component (As shown in Figure \ref{fig:spi_architecture}) integrates semantic information extracted from SAM with features derived from the IR model, following an approach inspired by the SAM-Deblur pipeline \cite{li2024sam}. 
This integration ensures efficient inference without compromising the performance of the original IR model while leveraging valuable semantic priors.

In the SPI process, the original IR model $f^{IR1}$ first processes the degraded image $I_{LQ}$ to produce an initial restored image $I_{HQ}^1$, as described by the equation $I_{HQ}^1 = f^{IR1}(I_{LQ})$.
Concurrently, we utilize the ground truth (GT) image $I_{HQ}$ to generate semantic masks $M$ that capture key anatomical structures. 
This is represented as $M = SAM(I_{HQ})$, where SAM is our fine-tuned Segment Anything Model. 
By using the GT image for mask generation, we ensure that our semantic priors are based on the most accurate anatomical information available.

The core of the SPI process lies in the Mask Average Pooling (MAP) Unit, which takes both $I_{HQ}^1$ and $M$ as inputs. 
Within this unit, an encoder embedded in the second IR model $f^{IR2}$ extracts features $F_{enc}$ from $I_{HQ}^1$. The MAP operation then computes the average of $F_{enc}$ within each region defined by the masks in $M$, reassigning these average values to produce $F_{MAP}$.

Following the MAP operation, a feature fusion step concatenates $F_{MAP}$ with $I_{HQ}^1$, creating the input $F_{input} = Concat[F_{MAP}, I_{HQ}^1]$ for the second IR model. 
This model, $f^{IR2}$, enhanced with the integrated semantic priors, then produces the final high-quality restored image $I_{HQ}^2 = f^{IR2}(F_{input})$.

This SPI approach ensures the effective incorporation of high-quality semantic priors into the restoration process, guiding it with valuable and accurate anatomical information specific to bone scintigraphy images. 
The use of GT images for mask generation during training provides our model with the most reliable semantic guidance possible. 
The use of two IR models ($f^{IR1}$ and $f^{IR2}$) allows for a two-stage restoration process, where the second stage is enhanced by precise semantic information derived from the ground truth.

\subsection{Semantic Knowledge Distillation}

To enhance the performance of the original IR model, we introduce the Semantic Knowledge Distillation component. SKD is designed to transfer the semantic understanding embedded in the enhanced IR model $f^{IR2}$ back to the original IR model $f^{IR1}$. This process is crucial for improving the performance of $f^{IR1}$ without explicitly incorporating semantic information during inference.

The concept of knowledge distillation, originally proposed by Hinton et al. \cite{hinton2015distilling}, involves transferring knowledge from a complex model to a simpler one. In our context, we adapt this concept to transfer semantic understanding. The key insight is that $f^{IR2}$, which has access to semantic priors, produces high-quality restored images that implicitly encode semantic information. By encouraging $f^{IR1}$ to mimic the output of $f^{IR2}$, we aim to transfer this implicit semantic understanding.

The second IR model $f^{IR2}$ produces a high-quality restored image $I_{HQ}^2$ by leveraging semantic priors. We define a distillation loss $\mathcal{L}_{SKD}$ to minimize the discrepancy between $I_{HQ}^1$ and $I_{HQ}^2$, encouraging the original IR model to learn from the enhanced model's output and improve its performance. The distillation loss is formulated as:

\begin{equation}
\mathcal{L}_{SKD} = 
\begin{cases}
|I_{HQ}^1 - I_{HQ}^2| - 0.5 & \text{if } |I_{HQ}^1 - I_{HQ}^2| > 1 ,\\
0.5 \times (I_{HQ}^1 - I_{HQ}^2)^2 & \text{if } |I_{HQ}^1 - I_{HQ}^2| \leq 1.
\end{cases}
\end{equation}

This piecewise loss function is designed to provide robustness to outliers while maintaining sensitivity to small differences. For large discrepancies ($>1$), it behaves like an L1 loss, providing resilience to extreme differences. For smaller discrepancies ($ \leq 1 $), it acts as an L2 loss, encouraging precise matching when the outputs are already close. The function is continuous at the transition point ($|I_{HQ}^1 - I_{HQ}^2| = 1$), ensuring smooth optimization.
The SKD process enables $f^{IR1}$ to implicitly learn semantic features without the need for explicit semantic information during inference. This is particularly valuable in the context of rapid bone scintigraphy image restoration, where the efficiency of the restoration process is crucial. By distilling knowledge from $f^{IR2}$, we aim to enhance $f^{IR1}$'s ability to produce high-quality restorations that respect the underlying anatomical structures, even when operating independently during inference.

\subsection{Semantic Consistency Module}

To enforce consistent semantic feature representation between the two restoration branches, we introduce a Semantic Consistency Module (SCM). Let the high-resolution outputs from the two restoration paths be \(X^A\) and \(X^B\). We first extract deep feature maps using a fixed, pretrained convolutional network \(\phi\) (VGG):

\begin{equation}
H^A = \phi\bigl(X^A\bigr), 
\quad
H^B = \phi\bigl(X^B\bigr),
\end{equation}
where \(H^A, H^B \in \mathbb{R}^{C \times H \times W}\) are feature tensors with \(C\) channels and spatial dimensions \(H \times W\).

Let \(\{S_k\}_{k=1}^K\) be \(K\) binary masks produced by a segmentation model (SAM), where each mask \(S_k \in \{0,1\}^{H \times W}\) isolates one semantic region. We apply these masks to the feature maps and then vectorize:

\begin{equation}
h^A_k = \mathrm{vec}\bigl(H^A \odot S_k\bigr), 
\quad
h^B_k = \mathrm{vec}\bigl(H^B \odot S_k\bigr),
\end{equation}
where \(\odot\) denotes element-wise multiplication and \(\mathrm{vec}(\cdot)\) flattens a tensor into a vector.

Next, for every pair of regions \((i,j)\), we compute the cosine similarity between their feature vectors to capture high-level semantic relationships:

\begin{equation}
c^A_{ij} = \frac{\langle h^A_i,\,h^A_j\rangle}{\|h^A_i\|_2\,\|h^A_j\|_2},
\quad
c^B_{ij} = \frac{\langle h^B_i,\,h^B_j\rangle}{\|h^B_i\|_2\,\|h^B_j\|_2}.
\end{equation}

Finally, we define the \emph{semantic alignment loss} to encourage the two branches to preserve the same relational structure among regions:

\begin{equation}
\mathcal{L}_{\mathrm{SCM}}
= \frac{1}{K(K-1)}
\sum_{\substack{i,j=1\\i\neq j}}^{K}
\bigl|\,c^A_{ij} - c^B_{ij}\bigr|.
\end{equation}

By integrating \(\mathcal{L}_{\mathrm{SCM}}\) into the overall training objective, the model not only achieves pixel-level accuracy but also maintains semantic consistency across restored images, leading to higher visual fidelity and improved performance on downstream tasks.

\subsection{Overall Optimization}

The overall loss function combines the reconstruction losses of $f^{IR1}$ models with the distillation and SCM losses:

\begin{equation}
\mathcal{L} = \mathcal{L}_{recon}^1 + \lambda_1 \mathcal{L}_{SKD} + \lambda_2 \mathcal{L}_{SCM},
\end{equation}
where $\lambda_1$ and $\lambda_2$ are weighting factors that balance the contributions of the distillation and SCM losses. 

For the reconstruction loss $\mathcal{L}_{recon}^1$, we employ the negative Peak Signal-to-Noise Ratio (PSNR) loss, commonly referred to as PSNR Loss. PSNR is a widely used metric in image processing to quantify the quality of reconstruction, particularly valuable in medical imaging where preserving fine details is crucial. It is defined as the ratio between the maximum possible signal power and the power of distorting noise that affects the fidelity of its representation. The PSNR Loss is computed as the negative of the PSNR value between the ground truth high-quality image $I_{HQ}$ and the output of our $f^{IR1}$ model:

\begin{equation}
\mathcal{L}_{recon}^1 = -PSNR(I_{HQ}, f^{IR1}(I_{LQ})).
\end{equation}

This loss encourages the model to produce reconstructed images that closely match the ground truth in terms of PSNR, which is essential for maintaining high fidelity in the restored medical images.

During the training process, we focus the optimization solely on the parameters of the $f^{IR1}$ model. The parameters of $f^{IR2}$, SAM, and the VGG network used in the SCM are kept frozen throughout the training. This strategic decision serves to maintain the stability of the semantic prior generation, ensure consistency in the knowledge distillation process, and preserve the reliability of the semantic consistency checks. By keeping these components unchanged, we effectively guide $f^{IR1}$ to learn from the semantically enhanced results of $f^{IR2}$ and the semantic consistency information provided by the SCM, without altering the semantic understanding embedded in these modules.

This approach ensures that semantic priors are effectively utilized during the training phase while avoiding any increase in computational burden during inference. Once training is complete, $f^{IR1}$ can operate independently, having implicitly learned to incorporate semantic information in its restoration process. This makes our method particularly suitable for rapid bone scintigraphy image restoration, where both high-quality results and computational efficiency are of paramount importance.

\subsection{Fine-Tuning Segment Anything Model (SAM)}

\begin{figure*}[!ht]
\setlength{\abovecaptionskip}{0pt}
\setlength{\belowcaptionskip}{0pt}
\centering
\includegraphics[scale=0.77]{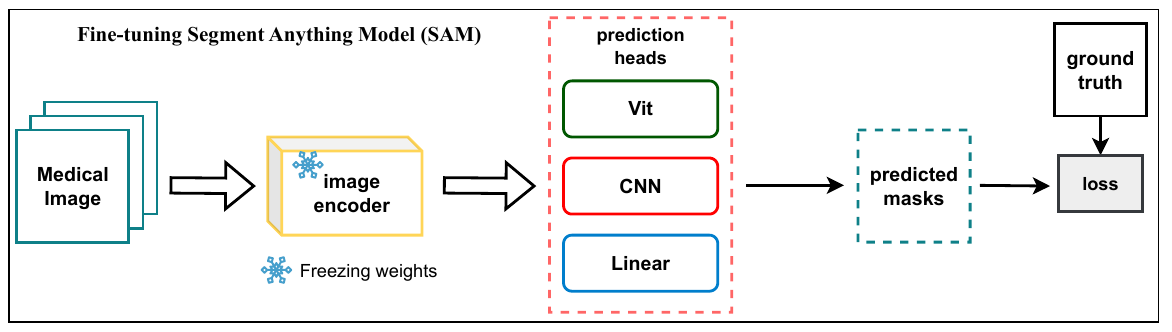}
\caption{Fine-tuning architecture of the Segment Anything Model (SAM) for image segmentation. The pre-trained image encoder weights are frozen, while a lightweight CNN prediction head is introduced and trained on task-specific data. The model takes medical images as input and outputs predicted segmentation masks, which are compared to ground truth labels to compute the loss for training.}
\vspace{-.1in}
\label{fig2}
\end{figure*}

To adapt SAM for our specific application, we employ a fine-tuning strategy inspired by AutoSAM \cite{hu2023efficiently}. 
This approach has shown promising results in adapting SAM for medical image segmentation (As shown in Figure \ref{fig2}). 
The fine-tuning process involves maintaining the pre-trained weights of the SAM encoder to leverage its general feature extraction capabilities. 
We introduce a lightweight, task-specific Convolutional Neural Network (CNN) head for training. 

This CNN head consists of a reshaping layer that transforms the image embedding into a feature map of size (256, 64, 64), followed by multiple stages of convolutional layers. 
Each stage includes a convolutional layer with a stride of 1 and a transposed convolutional layer with a stride of 2 for upscaling. A final point-wise convolutional layer with a kernel size of 1 produces prediction masks for each class.
This fine-tuning strategy enables efficient adaptation of SAM to the specific requirements of images while maintaining the benefits of its pre-trained features.

The training process utilized an optimization strategy tailored for stability and performance. Specifically, the AdamW optimizer was employed with an initial learning rate of 0.005. A warmup approach was implemented to stabilize the early training stages, where the learning rate was gradually increased over the first 250 iterations before transitioning to the main training phase. The fine-tuning process spanned a maximum of 60 epochs to ensure sufficient convergence without overfitting.
To guide the model's training, a composite loss function was applied, combining cross-entropy loss and Dice loss in a $0.2:0.8$ ratio.

\section{Experiments}

\subsection{Datasets}

\subsubsection{Rapid Bone Scintigraphy Dataset}

\begin{figure}[!ht]
\setlength{\abovecaptionskip}{0pt}
\setlength{\belowcaptionskip}{0pt}
\centering
\includegraphics[scale=0.38]{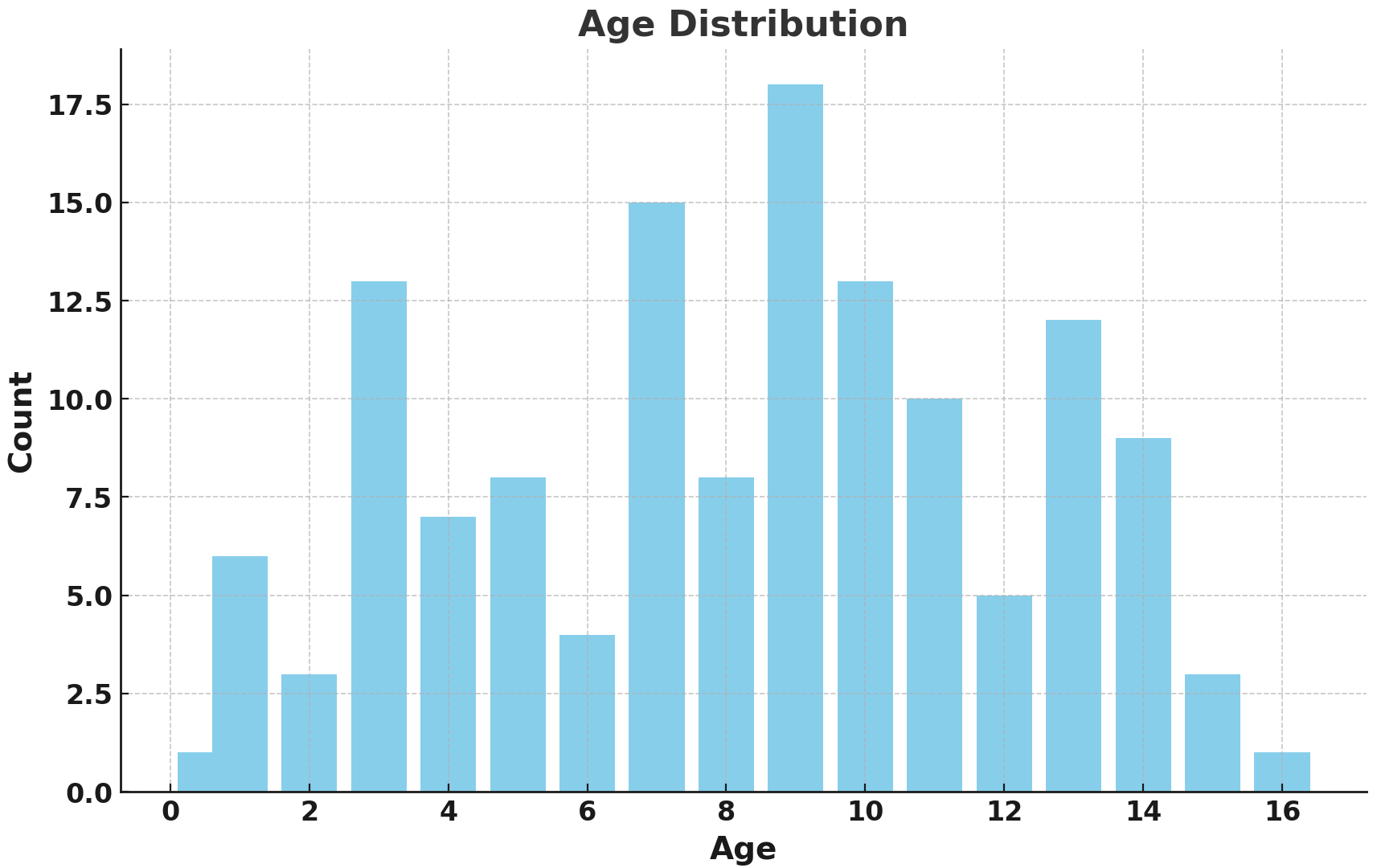}
\caption{Age distribution of patients in RBS dataset. The histogram shows the number of patients for each age group from 0.5 to 16 years old. The y-axis represents the number of patients, while the x-axis shows the age in years.}
\vspace{-.1in}
\label{fig:bone_scintigraphy_age}
\end{figure}

\begin{figure}[!ht]
\setlength{\abovecaptionskip}{0pt}
\setlength{\belowcaptionskip}{0pt}
\centering
\includegraphics[scale=0.07]{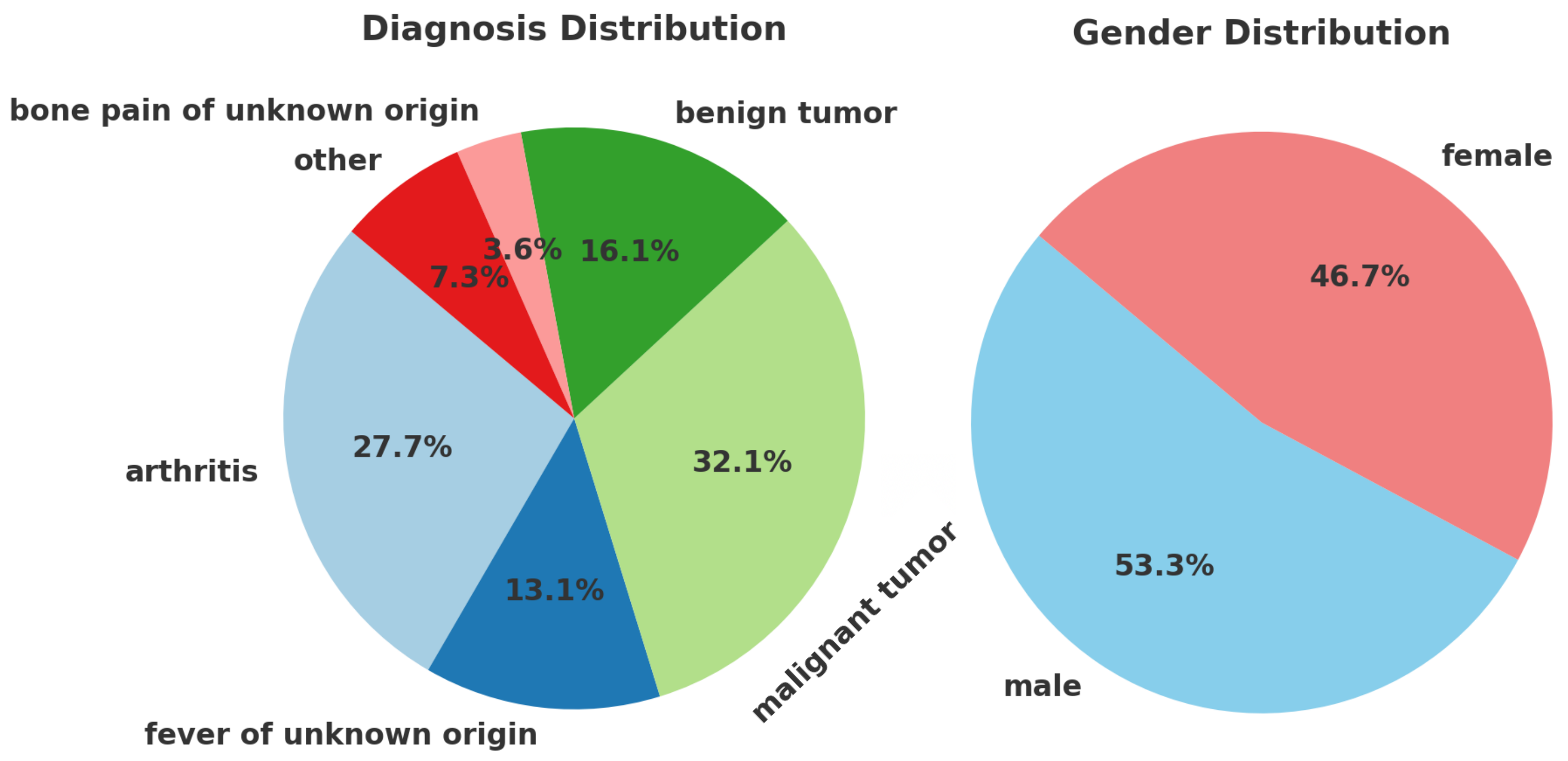}
\caption{Distribution of diagnoses and gender among the RBS dataset. The left pie chart illustrates the proportion of different diagnosis categories: arthritis, fever of unknown origin, malignant tumor, benign tumor, bone pain of unknown origin, and other. The right pie chart shows the gender distribution within the same population, highlighting the proportion of male and female subjects.}
\vspace{-.1in}
\label{fig:bone_scintigraphy_gender}
\end{figure}

Our primary experiments were conducted on a Rapid Bone Scintigraphy (RBS) dataset comprising 137 pediatric patients, aged 0.5 to 16 years (age distribution shown in Figure~\ref{fig:bone_scintigraphy_age}), who underwent both standard and rapid bone scintigraphy scans. 
The scans were performed using a SPECT system, specifically the GE Discovery NM/CT 670 Pro. The scanning speeds were 20 cm/min for standard scans and 40 cm/min for rapid scans, representing a 2× acceleration.
As shown in Figure \ref{fig:bone_scintigraphy_gender}, the study population consisted of 73 males and 64 females, with a mean age of 8.16 years (SD = 3.94 years). Data was collected retrospectively from January 2023 to January 2024.
The most common clinical diagnosis was a malignant tumor (n = 44, 32.1\%), followed by arthritis (n = 38, 27.7\%), and benign tumor (n = 22, 16.1\%). Other diagnoses included fever of unknown origin (n = 18, 13.1\%), bone pain of unknown origin (n = 5, 3.6\%), and other conditions (n = 10, 7.3\%). 
This diverse range of diagnoses provided a comprehensive representation of typical cases encountered in pediatric bone scintigraphy. This study received approval from the Ethics Committee of Shanghai Children's Hospital (Approval number: 2024R078).

\begin{figure}[!ht]
\setlength{\abovecaptionskip}{0pt}
\setlength{\belowcaptionskip}{0pt}
\centering
\includegraphics[scale=0.05]{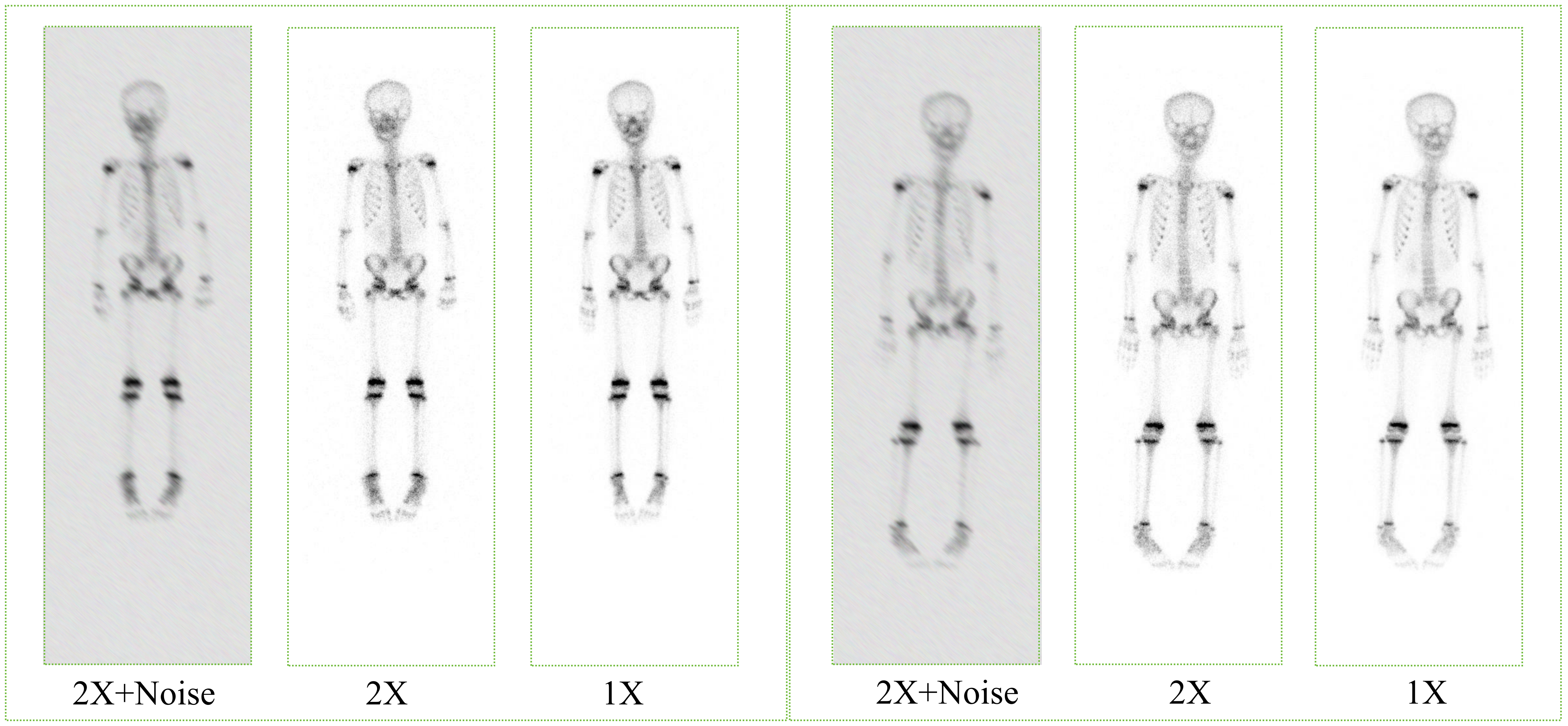}
\caption{Comparison of bone scintigraphy images. From left to right: 2X+Noise (rapid scan with added noise), 2X (rapid scan), and 1X (standard scan). The top and bottom rows show different patient cases.}
\vspace{-.1in}
\label{fig:bone_scintigraphy}
\end{figure}

To further simulate and address the challenges of rapid scans, we applied additional processing to the rapid scan images. We introduced Gaussian noise with a mean 0 and variance of 0.01, speckle noise with a mean 0 and variance of 0.01, and motion blur with a degree of 10 and an angle of 5 degrees. These additions enhanced the realism of the rapid scan images and ensured the robustness of our restoration method. Figure \ref{fig:bone_scintigraphy} presents a comparison of bone scintigraphy images from our dataset. 

To address positional misalignment between the rapid and standard scan images due to potential patient movement during consecutive scans, we employed a patch-based registration approach. This method aligned corresponding patches from the rapid and standard scan images to correct translational errors while preserving inherent noise patterns and structural details. The registration process for each patch is given by:

\begin{equation}
p_{i,j}^{f \mid \text{reg}} = \mathcal{R}(p_{i,j}^{s} \leftarrow p_{i,j}^{f}),
\end{equation}
where \( \mathcal{R}(\cdot) \) denotes the registration function that aligns the patches.
Following preprocessing, both the rapid and standard scan images were uniformly cropped into patches of size 192×192, with a step size of 192 and a threshold size of 0, as specified in the configuration options. This ensured a consistent patch size for model training, facilitating efficient and effective learning.

\subsubsection{Low-Quality Endoscopic Surgical Dataset}

To validate our method's versatility, we extended experiments to the Endoscapes2023 dataset \cite{murali2023endoscapes} of endoscopic surgical videos. We randomly selected 1117 images from various video segments to maintain diversity. Synthetic smoke was generated using Gaussian blur on random noise and blended with the original images to simulate low-quality conditions. The images were then processed using a patch-based cropping approach, uniformly cropping them into 192×192 patches with a step size of 192 and no threshold size. This extension to the endoscopic surgical domain allowed us to demonstrate our method's effectiveness across different medical imaging modalities, further validating its robustness in handling diverse medical imaging challenges.

\subsection{Evaluation Metrics}

To comprehensively assess our proposed method's performance, we employed a diverse set of evaluation metrics \cite{zhang2024distilling,pan2024fast}. These metrics were chosen to capture different aspects of image quality, including fidelity, perceptual similarity, and overall visual quality. 
We utilized four key metrics: Peak Signal-to-Noise Ratio (PSNR), Structural Similarity Index (SSIM), Fréchet Inception Distance (FID), and Learned Perceptual Image Patch Similarity (LPIPS). PSNR and SSIM measure the fidelity and structural similarity between restored and reference images, respectively. FID evaluates the overall quality of restored images by measuring the distance between feature distributions. LPIPS assesses perceptual similarity using deep features. These metrics provide a comprehensive evaluation of our image restoration method's effectiveness.

\subsection{Implementation Details}

Our framework, implemented using PyTorch on NVIDIA A6000 GPUs, employed the Adam optimizer ($\beta_1=0.9$, $\beta_2=0.999$) for all components. We adopted a two-stage training strategy: first pre-training the $f^{IR2}$ model with SAM semantic information, followed by a distillation stage connecting $f^{IR2}$ to $f^{IR1}$ and optimizing only $f^{IR1}$. 
For the bone scintigraphy dataset, each stage lasted 60 epochs, while the endoscopic dataset used 30 epochs per stage. We maintained a learning rate of 1e-4 and a batch size of 8 across both datasets, splitting them into 70\% training, 15\% validation, and 15\% testing sets. SAM was fine-tuned using 10\% of labeled data from each dataset. 
This approach leverages semantic priors during training while ensuring efficient inference by distilling knowledge from $f^{IR2}$ to $f^{IR1}$, eliminating the need for SAM during inference. Our strategy effectively balances semantic-guided restoration with efficient processing, crucial for clinical applications where speed is essential.

\subsection{Selected Baseline Methods}

To evaluate our proposed method's effectiveness, we selected four state-of-the-art image restoration models as baselines: EDSR \cite{lim2017enhanced}, Restormer \cite{zamir2022restormer}, SwinIR \cite{liang2021swinir}, and NAFNet \cite{chen2022simple}. These models represent diverse architectural approaches and have demonstrated strong performance in various image restoration tasks. EDSR features a deep residual structure with skip connections, allowing for effective gradient flow. 
Restormer leverages self-attention mechanisms to capture long-range dependencies, making it effective for complex degradations. SwinIR utilizes the Swin Transformer architecture, offering hierarchical representation learning with its shifted window approach. NAFNet achieves competitive performance without nonlinear activation functions, reducing computational complexity while maintaining high restoration quality. Each model brings unique strengths to medical image restoration, where both efficiency and accuracy are crucial.

\subsection{Quantitative and Qualitative Results}

\subsubsection{Pediatric Rapid Bone Scintigraphy Images}

The experimental results for pediatric rapid bone scintigraphy image restoration demonstrate the effectiveness of our proposed method across various state-of-the-art baseline models. 
Table \ref{tab:pediatric_results_comparison} summarizes the quantitative results, showcasing the performance of IR1 in our framework.

\begin{table*}[!ht]
\setlength{\abovecaptionskip}{0pt}
\setlength{\belowcaptionskip}{0pt}
\centering
\small
\caption{Quantitative results on the RBS dataset. The '+' symbol indicates models enhanced with our proposed semantic prior integration method. Arrows indicate improvement (↑) or decline (↓) compared to the base model.}
\label{tab:pediatric_results_comparison}
\begin{tabular}{lcccc}
\hline
Method & PSNR ↑ & SSIM ↑ & FID ↓ & LPIPS ↓ \\
\hline
Restormer & 37.35 & 0.9248 & 36.73 & 0.1133 \\
Restormer+ & \textbf{38.00} & 0.9238  & 37.70  & \textbf{0.1130} \\
\hline
SwinIR & 33.32 & 0.9163 & 41.19 & 0.1203 \\
SwinIR+ & \textbf{34.14} & \textbf{0.9163} & \textbf{40.63 } & 0.1217  \\
\hline
EDSR & 31.99 & 0.9167 & 35.27 & 0.1144 \\
EDSR+ & 31.93  & \textbf{0.9187 } & 35.37  & 0.1145  \\
\hline
RCAN & 33.40 & 0.9088 & 34.32 & 0.1137 \\
RCAN+ & \textbf{34.25 } & \textbf{0.9097 } & 35.39  & 0.1195  \\
\hline
NAFNet & 38.52 & 0.9223 & 30.59 & 0.1113 \\
NAFNet+ & \textbf{39.39} & \textbf{0.9227 } & \textbf{28.93 } & \textbf{0.1112 } \\
\hline
\vspace{-.1in}
\end{tabular}
\end{table*}

\begin{figure*}[!ht]
\setlength{\abovecaptionskip}{0pt}
\setlength{\belowcaptionskip}{0pt}
\centering
\includegraphics[scale=0.12]{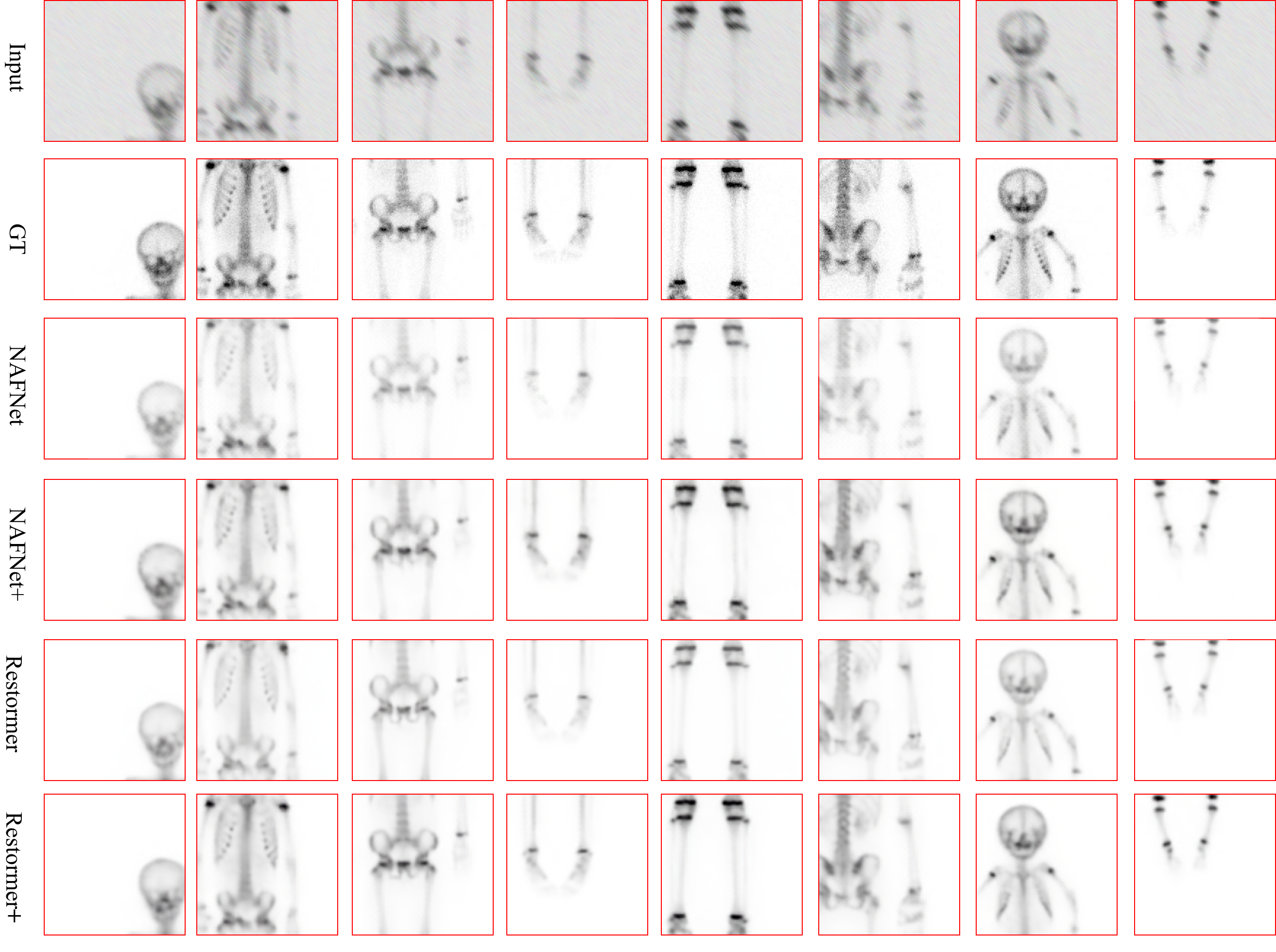}
\caption{Comparison of different restoration methods on bone scintigraphy image patches. From top to bottom: Input (noisy rapid scan), Ground Truth (GT), NAFNet, NAFNet+, Restormer, and Restormer+. Columns show different anatomical regions.}
\vspace{-.1in}
\label{fig:restoration_comparison}
\end{figure*}

Our method generally enhances the performance of existing models across various metrics. NAFNet+ achieved the highest PSNR of 39.39 dB, a significant improvement of 0.87 dB over the original NAFNet. SSIM scores remained stable or slightly improved for most models, with NAFNet+ showing the best performance (0.9227). FID scores varied across models, but NAFNet+ achieved the lowest (best) score of 28.93, indicating enhanced perceptual quality. LPIPS scores showed minor improvements or remained stable, with NAFNet+ achieving the lowest score of 0.1112.
Among tested models, NAFNet and its enhanced version (NAFNet+) consistently outperformed other approaches across all metrics. While some models (e.g., EDSR) showed mixed results, the overall trend indicates that our method generally improves or maintains the performance of existing state-of-the-art models.

Figure \ref{fig:restoration_comparison} provides a visual comparison of different restoration methods. NAFNet+ and Restormer+ demonstrate significant improvements across all anatomical regions compared to their baseline counterparts, exhibiting superior performance in noise reduction, contrast enhancement, and preservation of fine anatomical details. NAFNet+ consistently shows the most impressive results, closely resembling the ground truth images in most regions.

\subsubsection{Low-Quality Endoscopic Surgical Datasets}

Table \ref{tab:endoscopic_results_comparison} presents the quantitative results of our semantic prior integration method on the Low-Quality Endoscopic Surgical Dataset. Our method showed improvements in PSNR and SSIM for several models, with NAFNet+ achieving the highest PSNR of 39.43 dB (0.21 dB improvement) and SSIM of 0.9925. RCAN+ demonstrated substantial SSIM improvement from 0.9877 to 0.9910. FID and LPIPS scores showed mixed results, suggesting nuanced impacts on perceptual quality.

\begin{table*}[!ht]
\setlength{\abovecaptionskip}{0pt}
\setlength{\belowcaptionskip}{0pt}
\centering
\small
\caption{Quantitative results on the low-quality endoscopic surgical dataset. The '+' symbol indicates models enhanced with our proposed semantic prior integration method. Arrows indicate improvement (↑) or decline (↓) compared to the base model.}
\label{tab:endoscopic_results_comparison}
\begin{tabular}{lcccc}
\hline
Method & PSNR ↑ & SSIM ↑ & FID ↓ & LPIPS ↓ \\
\hline
Restormer & 39.17 & 0.9915 & 1.65 & 0.0113 \\
Restormer+ & \textbf{39.28 } & \textbf{0.9919 } & 1.73 & 0.0119  \\
\hline
SwinIR & 39.06 & 0.9907 & 1.69 & 0.0130 \\
SwinIR+ & 38.78  & 0.9904  & \textbf{1.68 } & \textbf{0.0125} \\
\hline
EDSR & 38.81 & 0.9899 & 1.70 & 0.0129 \\
EDSR+ & 38.80  & 0.9899 & \textbf{1.65 } & 0.0125  \\
\hline
RCAN & 38.93 & 0.9877 & 1.62 & 0.0119 \\
RCAN+ & \textbf{39.28} & \textbf{0.9910} & 1.67  & 0.0119 \\
\hline
NAFNet & 39.22 & 0.9922 & 1.63 & 0.0112 \\
NAFNet+ & \textbf{39.43} & \textbf{0.9925} & 1.68 & 0.0116  \\
\hline
\vspace{-.1in}
\end{tabular}
\end{table*}

\begin{figure}[!ht]
\setlength{\abovecaptionskip}{0pt}
\setlength{\belowcaptionskip}{0pt}
\centering
\includegraphics[scale=0.12]{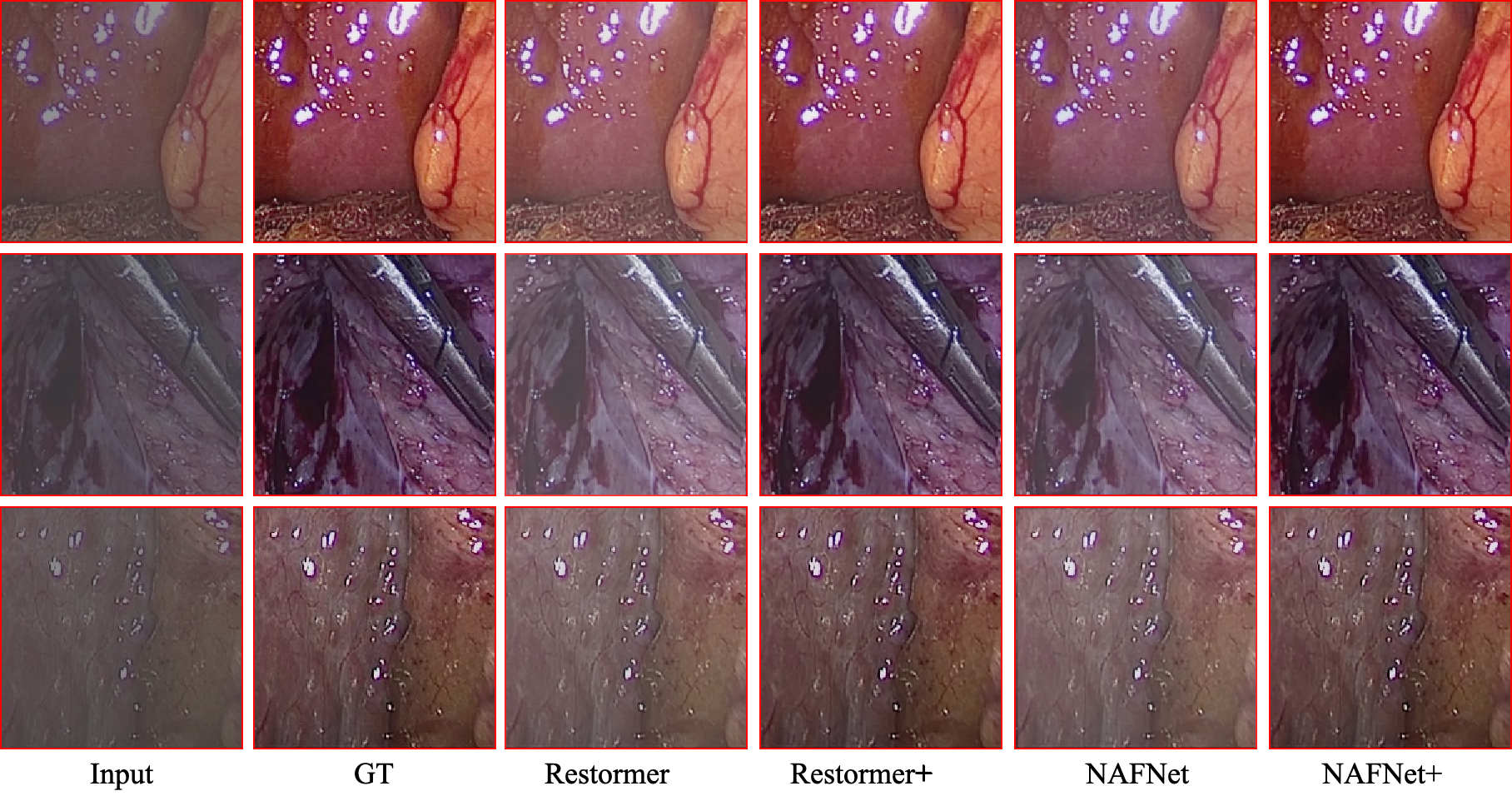}
\caption{Comparison of image restoration results on endoscopic surgical images. From left to right: Input (simulated low-quality image), Ground Truth (GT), Restormer, Restormer+, NAFNet, and NAFNet+. Each row represents a different endoscopic scene.}
\vspace{-.1in}
\label{fig:restoration_comparison1}
\end{figure}

Figure \ref{fig:restoration_comparison1} compares restoration results on three endoscopic scenes. NAFNet+ and Restormer+ showed significant improvements over their base versions, particularly in color accuracy, edge definition, and fine detail preservation. Overall, NAFNet+ exhibited a slight advantage, demonstrating the effectiveness of our method in enhancing low-quality endoscopic surgical image restoration.

\subsection{Ablation Study}

\begin{table*}[!ht]
\setlength{\abovecaptionskip}{0pt}
\setlength{\belowcaptionskip}{0pt}
\centering
\small
\caption{Ablation study results on the RBS dataset. Arrows indicate improvement (↑) or decline (↓) compared to the base model. Bold numbers indicate the best result for each metric.}
\label{tab:ablation_bone_comparison}
\begin{tabular}{lcccc}
\hline
Method & PSNR ↑ & SSIM ↑ & FID ↓ & LPIPS ↓ \\
\hline
$f^{IR1}$ (NAFNet) & 38.52 & 0.9223 & 30.59 & 0.1113 \\
+SKD, (w. SAM) & 38.92  & 0.9223  & 30.36  & 0.1124  \\
+SKD+SCM, (w. SAM) & \textbf{39.39 } & \textbf{0.9227 } & \textbf{28.93} & \textbf{0.1112} \\
\hline
\end{tabular}
\vspace{-.1in}
\end{table*}

To evaluate the effectiveness and progressive impact of different components in our proposed method, we conducted an ablation study on both the pediatric rapid bone scintigraphy and low-quality endoscopic surgical dataset. We compared the performance of the baseline model ($f^{IR1}$, using NAFNet as the backbone) with two progressive variants of our method: first incorporating Semantic Knowledge Distillation, and then adding the Semantic Consistency Module on top of SKD. Both variants utilized the SAM for semantic information. 
Tables \ref{tab:ablation_bone_comparison} and \ref{tab:ablation_endoscopic_comparison} present the results of this study.

The ablation study reveals the progressive improvements brought by each component in our proposed method. Semantic Knowledge Distillation: The addition of SKD to the baseline model ($f^{IR1}$) shows initial improvements in both datasets. For the pediatric rapid bone scintigraphy dataset, SKD increases the PSNR by 0.40 dB and reduces the FID by 0.23. In the Low-Quality Endoscopic Surgical dataset, SKD improves the PSNR by 0.08 dB. 
These results demonstrate that SKD effectively transfers semantic knowledge from the teacher model to the student model, enhancing the overall restoration quality. Semantic Consistency Module, Building upon the SKD, the integration of SCM further improves the performance across both datasets. 

In the RBS dataset, the combination of SKD and SCM achieves the best results in all metrics, with a significant PSNR improvement of 0.87 dB over the baseline and a substantial FID reduction of 1.66. For the Low-Quality Endoscopic Surgical dataset, the full model (SKD+SCM) also shows the highest PSNR (39.43 dB) and SSIM (0.9925) scores. These results highlight the effectiveness of SCM in maintaining semantic consistency throughout the restoration process, leading to improved image quality and structural similarity.

\begin{table*}[!ht]
\setlength{\abovecaptionskip}{0pt}
\setlength{\belowcaptionskip}{0pt}
\centering
\small
\caption{Ablation study results on the low-quality endoscopic surgical dataset. Arrows indicate improvement (↑) or decline (↓) compared to the base model. Bold numbers indicate the best result for each metric.}
\label{tab:ablation_endoscopic_comparison}
\begin{tabular}{lcccc}
\hline
Method & PSNR ↑ & SSIM ↑ & FID ↓ & LPIPS ↓ \\
\hline
$f^{IR1}$ (NAFNet) & 39.22 & 0.9922 & 1.63 & 0.0112 \\
+SKD, (w. SAM) & 39.30  & 0.9903  & 1.70 & 0.0117  \\
+SKD+SCM, (w. SAM) & 39.43  & 0.9925  & 1.68 & 0.0116  \\
\hline
\end{tabular}
\vspace{-.1in}
\end{table*}


\begin{table}[!ht]
\setlength{\abovecaptionskip}{0pt}
\setlength{\belowcaptionskip}{0pt}
\centering
\small
\caption{Impact of $\lambda_1$ and $\lambda_2$ on PSNR and SSIM. Bold numbers indicate the best result for each parameter and metric.}
\label{tab:combined_lambda_tuning}
\begin{tabular}{cccc}
\hline
Parameter & Value & PSNR ↑ & SSIM ↑ \\
\hline
{$\lambda_1$} & 0.0001 & 39.17 & \textbf{0.9227} \\
 & 0.001 & 39.24 & 0.9226 \\
 & 0.01 & \textbf{39.39} & \textbf{0.9227} \\
\hline
{$\lambda_2$} & 10 & 39.35 & 0.9226 \\
 & 100 & \textbf{39.39} & \textbf{0.9227} \\
 & 1000 & 39.31 & 0.9226 \\
\hline
\end{tabular}
\vspace{-.1in}
\end{table}

\begin{figure}[!ht]
\setlength{\abovecaptionskip}{0pt}
\setlength{\belowcaptionskip}{0pt}
\centering
\includegraphics[scale=0.36]{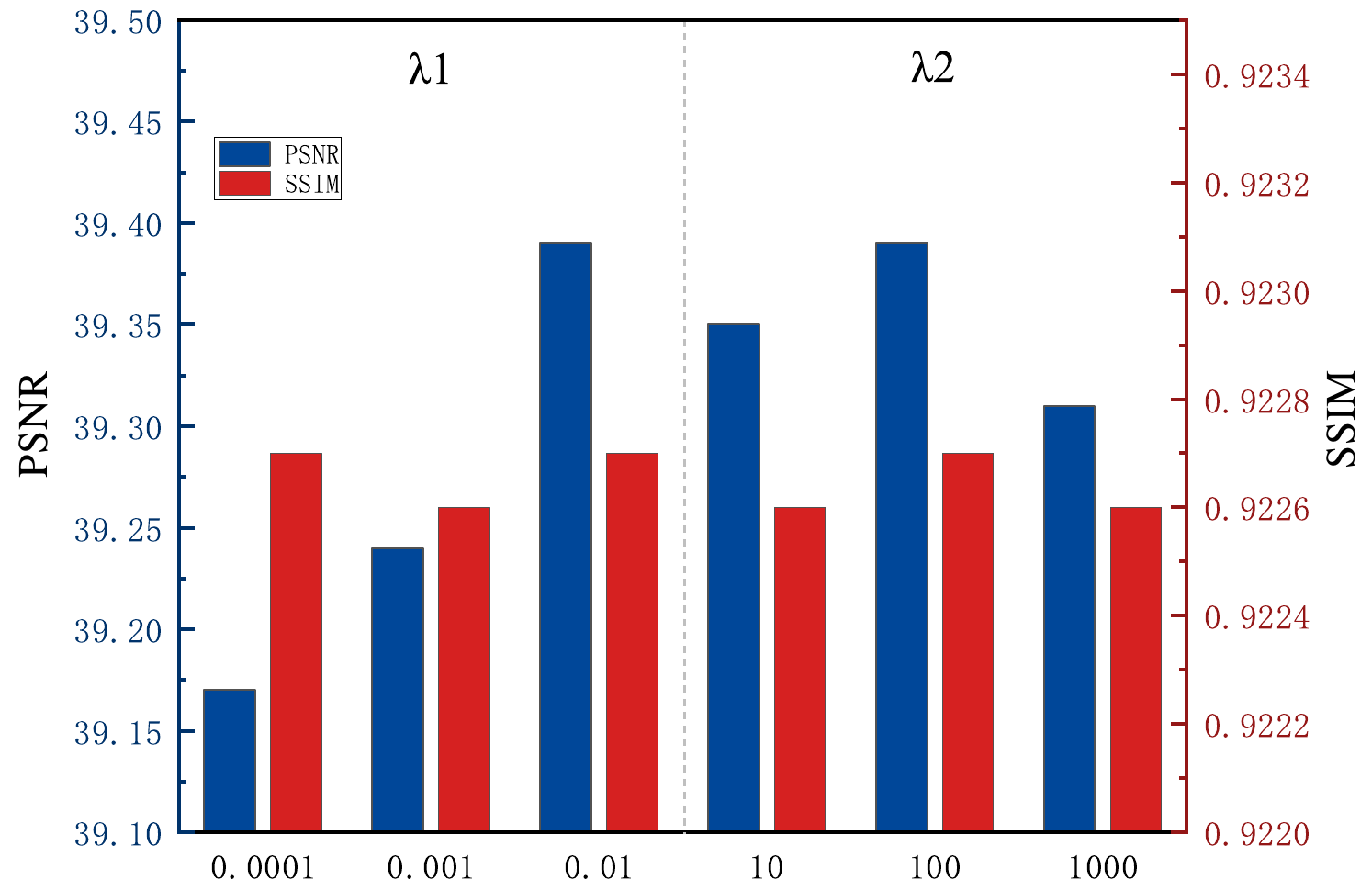}
\caption{Sensitivity analysis for $\lambda_1$ (SKD weight) and $\lambda_2$ (SCM weight) on the RBS dataset.}
\vspace{-.1in}
\label{fig6}
\end{figure}

We further investigated the sensitivity of our model to two key hyperparameters: $\lambda_1$, which controls the weight of the SKD loss, and $\lambda_2$, which governs the influence of the SCM. Table \ref{tab:combined_lambda_tuning} and Figure \ref{fig6} show these results.
The hyperparameter sensitivity analysis reveals that increasing $\lambda_1$ from 0.0001 to 0.01 led to a consistent improvement in PSNR, with the best performance at $\lambda_1 = 0.01$. SSIM remained relatively stable across different values, with a slight dip at 0.001. For $\lambda_2$, the model showed optimal performance at $\lambda_2 = 100$ for both PSNR and SSIM. Lower ($\lambda_2 = 10$) or higher ($\lambda_2 = 1000$) values led to slight decreases in performance, indicating the importance of properly balancing the influence of the Semantic Consistency Module.
These results demonstrate that our model's performance is sensitive to the choice of hyperparameters, particularly $\lambda_1$ and $\lambda_2$. The optimal values ($\lambda_1 = 0.01$ and $\lambda_2 = 100$) strike a balance between the influence of semantic knowledge distillation and semantic consistency preservation.

\subsection{Comparison of existing SAM priors introduction methods}

\begin{table*}[!ht]
\setlength{\abovecaptionskip}{0pt}
\setlength{\belowcaptionskip}{0pt}
\centering
\small
\caption{Comparison of different SAM priors introduction methods on the RBS dataset. Arrows indicate improvement (↑) or decline (↓) compared to the baseline.}
\label{tab:sam_priors_comparison}
\begin{tabular}{lcccc}
\hline
Method & PSNR ↑ & SSIM ↑ & FID ↓ & LPIPS ↓ \\
\hline
NAFNet (baseline) & 38.52 & 0.9223 & 30.59 & 0.1113 \\
\hline
+CAT, (w. SAM) & 37.09  & 0.9224  & 29.54  & 0.1132  \\
+SPI, (w. SAM) & \textbf{39.60 } & \textbf{0.9236 } & \textbf{30.24 } & 0.1126  \\
+SPI - no dropout, (w. SAM) & 39.36  & 0.9222  & 30.25  & 0.1144  \\
\hline
\end{tabular}
\vspace{-.1in}
\end{table*}

To evaluate different methods for introducing SAM priors into our $f^{IR2}$ model, we compared SPI with the Concatenation (CAT) method \cite{jin2023let} and a variant of SPI without dropout. 
Table \ref{tab:sam_priors_comparison} presents the results on the RBS dataset.
The CAT method, which directly concatenates SAM features with original image features, led to a significant drop in PSNR (37.09 dB) compared to the baseline (38.52 dB), despite showing slight improvements in other metrics.

In contrast, the SPI method demonstrates superior performance across all metrics. PSNR increased significantly to 39.60 dB, an improvement of 1.08 dB over the baseline. SSIM improved to 0.9236, while both FID and LPIPS decreased, indicating better perceptual quality and structural similarity.
The SPI variant without dropout shows good performance (39.36 dB PSNR), but is slightly lower than the full SPI method, underscoring the importance of dropout as a regularization technique in our model.
These results suggest that the SPI method can make better use of the semantic information provided by SAM, outperforming both the CAT method and the baseline in medical image restoration tasks.

\subsection{Fine-tuning SAM on Medical Datasets}

Although the Segment Anything Model (SAM) exhibits strong zero-shot segmentation performance on natural images, direct application to specialized medical scenarios may suffer from a large domain gap. 
To address this issue, we fine-tune SAM on a \textbf{small set of labeled medical images} in two distinct tasks: 
(1) pediatric bone scintigraphy segmentation, and 
(2) endoscopic surgical scene segmentation. 
This step is crucial to ensure that SAM produces reliable segmentation masks, which in turn guide our image restoration network effectively.

\subsubsection{Pediatric Bone Scintigraphy Segmentation Results}

We use 10 annotated bone scintigraphy images covering major anatomical regions, including skull, ribs, shoulder, spine, etc. 
For comparison, we train two baseline methods: 
\emph{U-Net} and \emph{ViT-tiny}, on the same 10-image subset. 
Table~\ref{tab:BoneSegDice} reports the \emph{Dice similarity coefficient} (Dice, \%) and \emph{Hausdorff Distance} (HD, mm) for each anatomical region, as well as the mean across all regions. 
As can be seen, our \textbf{fine-tuned SAM} achieves competitive or superior results compared to the baselines, especially in terms of segmentation accuracy (Dice).

\begin{table*}[!t]
\centering
\small
\caption{Segmentation performance on our pediatric bone scintigraphy dataset using 10 training images. We report Dice (\%) and Hausdorff Distance (HD, mm) for each anatomical region. The best results in each column are in \textbf{bold}.}
\label{tab:BoneSegDice}
\scalebox{0.607}{
\begin{tabular}{ccccccccccccccccc}
\hline
{Methods} & \multicolumn{2}{c}{Skull} & \multicolumn{2}{c}{Ribs} & \multicolumn{2}{c}{Shoulder joint} & \multicolumn{2}{c}{Spine} & \multicolumn{2}{c}{Upper Extremity} & \multicolumn{2}{c}{Lower Extremity} & \multicolumn{2}{c}{Pelvis} & \multicolumn{2}{c}{\textbf{Mean}}     \\ \cline{2-17} 
                         & Dice (\%)    & HD (mm)    & Dice (\%)    & HD (mm)   & Dice (\%)         & HD (mm)        & Dice (\%)    & HD (mm)    & Dice (\%)         & HD (mm)         & Dice (\%)         & HD (mm)         & Dice (\%)     & HD (mm)    & \textbf{Dice (\%)} & \textbf{HD (mm)} \\ \hline
UNet                     & 91.11        & 4.75       & 46.65        & 39.82     & 73.08             & 7.20           & 74.71        & 4.85       & 66.90             & 18.07           & 81.71             & 3.00            & 57.67         & 58.23      & 70.26              & 19.42            \\
VIT-tiny                 & 13.00        & 51.79      & 2.53         & 31.99     & 5.11              & 144.50         & 15.56        & 36.43      & 2.06              & 48.42           & 6.10              & 7.63            & 14.50         & 52.80      & 8.41               & 53.37            \\
\textbf{Fine-tune SAM}   & 89.39        & 2.47       & 56.70        & 16.79     & 78.43             & 7.24           & 82.02        & 3.37       & 69.73             & 6.89            & 74.51             & 3.91            & 77.37         & 5.57       & \textbf{75.45}     & \textbf{6.61}    \\ \hline
\end{tabular}}
\end{table*}

\begin{table}[!t]
\centering
\small
\caption{Segmentation performance on the endoscopic surgical dataset using 100 training images. We report the Dice similarity coefficient (\%) and Hausdorff Distance (HD, mm) for two label categories and the mean}
\label{tab:EndoSegDice}
\scalebox{0.67}{
\begin{tabular}{ccccccc}
\hline
{Methods} & \multicolumn{2}{c}{Organ tissue} & \multicolumn{2}{c}{Surgical instruments} & \multicolumn{2}{c}{\textbf{Mean}}     \\ \cline{2-7} 
                         & Dice (\%)        & HD (mm)       & Dice (\%)            & HD (mm)           & \textbf{Dice (\%)} & \textbf{HD (mm)} \\ \hline
UNet                     & 63.24            & 38.33         & 67.94                & 30.76             & 65.59              & 34.54            \\
VIT-tiny                 & 40.44            & 12.59         & 58.91                & 17.18             & 49.67              & 14.88            \\
\textbf{Fine-tune SAM}   & 69.30            & 16.70         & 69.34                & 16.85             & \textbf{69.32}     & \textbf{16.78}   \\ \hline
\end{tabular}}
\vspace{-12pt}
\end{table}

\subsubsection{Endoscopic Surgical Scene Segmentation Results}

We further validate our approach on an \textbf{endoscopic surgical} dataset. 
Following the same principle, we select 100 labeled images (covering organ tissues and surgical instruments) and compare our fine-tuned SAM with two baselines: U-Net and ViT-tiny. 
Table~\ref{tab:EndoSegDice} reports the segmentation accuracy (Dice) and HD. 
We observe that \textbf{fine-tuned SAM} again provides better or on-par performance, which indicates the \emph{effectiveness} of adapting large-scale foundation models to specific medical tasks with minimal labeled data.

The results in Tables~\ref{tab:BoneSegDice}--\ref{tab:EndoSegDice} confirm that our fine-tuned SAM significantly outperforms its direct zero-shot usage on medical images (which yields many incorrect labels and is hence omitted). 
These experiments validate that \textbf{fine-tuning a small segmentation head} is sufficient to adapt SAM to specialized medical domains. 
Hence, the reliability of SAM priors is improved, avoiding bottlenecks in our rapid bone scintigraphy restoration or endoscopic image enhancement tasks.

\section{Discussion}

This study demonstrates the significant clinical potential of our semantic-guided framework for enhancing rapid bone scintigraphy images in pediatric patients. The substantial improvements in image quality, as evidenced by enhanced PSNR, SSIM, FID, and LPIPS metrics, suggest important implications for diagnostic accuracy in clinical settings. The ability to visualize fine anatomical details and subtle bone abnormalities from rapid scans could lead to earlier detection of pathologies and potentially reduce the need for follow-up imaging, thereby streamlining the diagnostic process and minimizing radiation exposure.

Our method's capacity to maintain image quality while significantly reducing scan times (from 20 cm/min to 40 cm/min) addresses a critical challenge in pediatric nuclear medicine. Shorter scan durations can dramatically improve the patient experience, particularly for young or anxious children who may struggle with prolonged immobility. This improvement could potentially reduce the need for sedation, further enhancing patient safety and comfort.
From an operational perspective, the ability to produce high-quality diagnostic images from rapid scans could lead to significant improvements in clinical workflow and resource utilization. Shorter scan times allow for higher patient throughput, potentially reducing waiting times and improving access to this crucial diagnostic tool. The reduced likelihood of needing repeat scans due to poor image quality could also lead to more efficient use of imaging equipment and clinical staff time.

The successful application of our method to low-quality endoscopic surgical images demonstrates its potential versatility across different medical imaging modalities. This broad applicability suggests that our semantic-guided approach could have wide-ranging impacts beyond bone scintigraphy, potentially improving image quality in various diagnostic imaging procedures.


\section{Conclusion}

This study introduces a novel semantic-guided framework for enhancing rapid bone scintigraphy images in pediatric patients. Our dual-stage model, incorporating Semantic Prior Integration, Semantic Knowledge Distillation, and a Semantic Consistency Module, demonstrates significant improvements in image restoration.
The framework leverages SAM priors during training while maintaining computational efficiency during inference. By pre-training $f^{IR2}$ with semantic information and distilling this knowledge to $f^{IR1}$, we achieve a balance between semantic-guided restoration and efficient clinical application.
Experimental results on pediatric rapid bone scintigraphy images show consistent performance enhancements, with extensions to low-quality endoscopic surgical images demonstrating versatility across medical imaging modalities. Our comparative study reveals the superiority of our SPI approach over simpler concatenation methods, highlighting the benefits of sophisticated semantic information integration in medical image restoration tasks.

\bibliographystyle{IEEEtran}
\small
\bibliography{reference}

\begin{thebibliography}{10}
\providecommand{\url}[1]{#1}
\csname url@samestyle\endcsname
\providecommand{\newblock}{\relax}
\providecommand{\bibinfo}[2]{#2}
\providecommand{\BIBentrySTDinterwordspacing}{\spaceskip=0pt\relax}
\providecommand{\BIBentryALTinterwordstretchfactor}{4}
\providecommand{\BIBentryALTinterwordspacing}{\spaceskip=\fontdimen2\font plus
\BIBentryALTinterwordstretchfactor\fontdimen3\font minus \fontdimen4\font\relax}
\providecommand{\BIBforeignlanguage}[2]{{%
\expandafter\ifx\csname l@#1\endcsname\relax
\typeout{** WARNING: IEEEtran.bst: No hyphenation pattern has been}%
\typeout{** loaded for the language `#1'. Using the pattern for}%
\typeout{** the default language instead.}%
\else
\language=\csname l@#1\endcsname
\fi
#2}}
\providecommand{\BIBdecl}{\relax}
\BIBdecl

\bibitem{murata2024verification}
T.~Murata, T.~Hashimoto, M.~Onoguchi, T.~Shibutani, T.~Iimori, K.~Sawada, T.~Umezawa, Y.~Masuda, and T.~Uno, ``Verification of image quality improvement of low-count bone scintigraphy using deep learning,'' \emph{Radiological Physics and Technology}, vol.~17, no.~1, pp. 269--279, 2024.

\bibitem{isherwood2023sub}
A.~C. Isherwood, R.~Cabral, and G.~Avery, ``Sub 4 minute superfast single-photon emission computed tomography/computed tomography as an add-on for problem-solving in planar bone scintigraphy: a time-saving solution for departments without whole-body single-photon emission computed tomography/computed tomography,'' \emph{Nuclear Medicine Communications}, vol.~44, no.~5, pp. 407--413, 2023.

\bibitem{pan2024fast}
Z.~Pan, N.~Qi, Q.~Meng, B.~Pan, T.~Feng, J.~Zhao, and N.-J. Gong, ``Fast spect/ct planar bone imaging enabled by deep learning enhancement,'' \emph{Medical Physics}, 2024.

\bibitem{moncayo2021can}
V.~M. Moncayo, S.~Chimafor, E.~Lulaj, J.~A. Malko, and R.~Halkar, ``Can the diagnostic accuracy of bone scintigraphy be maintained with half the scanning time?'' \emph{Journal of Nuclear Medicine Technology}, vol.~49, no.~4, pp. 330--333, 2021.

\bibitem{alqahtani2023optimising}
M.~M.~M. Alqahtani, ``Optimising imaging protocols for whole-body spect/ct,'' Ph.D. dissertation, 2023.

\bibitem{ito2022adapting}
T.~Ito, T.~Maeno, H.~Tsuchikame, M.~Shishido, K.~Nishi, S.~Kojima, T.~Hayashi, and K.~Suzuki, ``Adapting a low-count acquisition of the bone scintigraphy using deep denoising super-resolution convolutional neural network,'' \emph{Physica Medica}, vol. 100, pp. 18--25, 2022.

\bibitem{pan2022ultra}
B.~Pan, N.~Qi, Q.~Meng, J.~Wang, S.~Peng, C.~Qi, N.-J. Gong, and J.~Zhao, ``Ultra high speed spect bone imaging enabled by a deep learning enhancement method: a proof of concept,'' \emph{EJNMMI physics}, vol.~9, no.~1, p.~43, 2022.

\bibitem{bahloul2024ultra}
A.~Bahloul, A.~Verger, Y.~Lamash, N.~Roth, D.~Dari, P.-Y. Marie, and L.~Imbert, ``Ultra-fast whole-body bone tomoscintigraphies achieved with a high-sensitivity 360° czt camera and a dedicated deep-learning noise reduction algorithm,'' \emph{European Journal of Nuclear Medicine and Molecular Imaging}, vol.~51, no.~5, pp. 1215--1220, 2024.

\bibitem{dong2015image}
C.~Dong, C.~C. Loy, K.~He, and X.~Tang, ``Image super-resolution using deep convolutional networks,'' \emph{IEEE transactions on pattern analysis and machine intelligence}, vol.~38, no.~2, pp. 295--307, 2015.

\bibitem{zhang2017beyond}
K.~Zhang, W.~Zuo, Y.~Chen, D.~Meng, and L.~Zhang, ``Beyond a gaussian denoiser: Residual learning of deep cnn for image denoising,'' \emph{IEEE transactions on image processing}, vol.~26, no.~7, pp. 3142--3155, 2017.

\bibitem{chen2017low}
H.~Chen, Y.~Zhang, M.~K. Kalra, F.~Lin, Y.~Chen, P.~Liao, J.~Zhou, and G.~Wang, ``Low-dose ct with a residual encoder-decoder convolutional neural network,'' \emph{IEEE transactions on medical imaging}, vol.~36, no.~12, pp. 2524--2535, 2017.

\bibitem{qi2023deep}
N.~Qi, B.~Pan, Q.~Meng, Y.~Yang, T.~Feng, H.~Liu, N.-J. Gong, and J.~Zhao, ``Deep learning enhanced ultra-fast spect/ct bone scan in patients with suspected malignancy: quantitative assessment and clinical performance,'' \emph{Physics in Medicine \& Biology}, vol.~68, no.~13, p. 135012, 2023.

\bibitem{jiang2023restore}
J.~Jiang and C.~Holz, ``Restore anything pipeline: Segment anything meets image restoration,'' \emph{arXiv preprint arXiv:2305.13093}, 2023.

\bibitem{xiao2023dive}
Z.~Xiao, J.~Bai, Z.~Lu, and Z.~Xiong, ``A dive into sam prior in image restoration,'' \emph{arXiv preprint arXiv:2305.13620}, 2023.

\bibitem{huang2022segmentation}
Z.~Huang, Z.~Liu, P.~He, Y.~Ren, S.~Li, Y.~Lei, D.~Luo, D.~Liang, D.~Shao, Z.~Hu \emph{et~al.}, ``Segmentation-guided denoising network for low-dose ct imaging,'' \emph{Computer Methods and Programs in Biomedicine}, vol. 227, p. 107199, 2022.

\bibitem{yin2022segmentation}
Z.~Yin and Z.~Zheng, ``Segmentation as domain knowledge in gan for low-dose ct denoising.'' \emph{Journal of Imaging Science \& Technology}, vol.~66, no.~4, 2022.

\bibitem{kirillov2023segment}
A.~Kirillov, E.~Mintun, N.~Ravi, H.~Mao, C.~Rolland, L.~Gustafson, T.~Xiao, S.~Whitehead, A.~C. Berg, W.-Y. Lo \emph{et~al.}, ``Segment anything,'' in \emph{Proceedings of the IEEE/CVF International Conference on Computer Vision}, 2023, pp. 4015--4026.

\bibitem{wu2023medical}
J.~Wu, W.~Ji, Y.~Liu, H.~Fu, M.~Xu, Y.~Xu, and Y.~Jin, ``Medical sam adapter: Adapting segment anything model for medical image segmentation,'' \emph{arXiv preprint arXiv:2304.12620}, 2023.

\bibitem{hu2023efficiently}
X.~Hu, X.~Xu, and Y.~Shi, ``How to efficiently adapt large segmentation model (sam) to medical images,'' \emph{arXiv preprint arXiv:2306.13731}, 2023.

\bibitem{li2024adapting}
K.~Li and P.~Rajpurkar, ``Adapting segment anything models to medical imaging via fine-tuning without domain pretraining,'' in \emph{AAAI 2024 Spring Symposium on Clinical Foundation Models}, 2024.

\bibitem{cheng2024unleashing}
Z.~Cheng, Q.~Wei, H.~Zhu, Y.~Wang, L.~Qu, W.~Shao, and Y.~Zhou, ``Unleashing the potential of sam for medical adaptation via hierarchical decoding,'' in \emph{Proceedings of the IEEE/CVF Conference on Computer Vision and Pattern Recognition}, 2024, pp. 3511--3522.

\bibitem{zhang2024sam}
Z.~Zhang, B.~Wang, W.~Liang, Y.~Li, X.~Guo, G.~Wang, S.~Li, and G.~Wang, ``Sam-guided enhanced fine-grained encoding with mixed semantic learning for medical image captioning,'' in \emph{ICASSP 2024-2024 IEEE International Conference on Acoustics, Speech and Signal Processing (ICASSP)}.\hskip 1em plus 0.5em minus 0.4em\relax IEEE, 2024, pp. 1731--1735.

\bibitem{sagheer2020review}
S.~V.~M. Sagheer and S.~N. George, ``A review on medical image denoising algorithms,'' \emph{Biomedical signal processing and control}, vol.~61, p. 102036, 2020.

\bibitem{goyal2018noise}
B.~Goyal, S.~Agrawal, and B.~Sohi, ``Noise issues prevailing in various types of medical images,'' \emph{Biomedical \& Pharmacology Journal}, vol.~11, no.~3, p. 1227, 2018.

\bibitem{florkow2022magnetic}
M.~C. Florkow, K.~Willemsen, V.~V. Mascarenhas, E.~H. Oei, M.~van Stralen, and P.~R. Seevinck, ``Magnetic resonance imaging versus computed tomography for three-dimensional bone imaging of musculoskeletal pathologies: a review,'' \emph{Journal of Magnetic Resonance Imaging}, vol.~56, no.~1, pp. 11--34, 2022.

\bibitem{katsaggelos2012digital}
A.~K. Katsaggelos, \emph{Digital image restoration}.\hskip 1em plus 0.5em minus 0.4em\relax Springer Publishing Company, Incorporated, 2012.

\bibitem{yang2024all}
Z.~Yang, H.~Chen, Z.~Qian, Y.~Yi, H.~Zhang, D.~Zhao, B.~Wei, and Y.~Xu, ``All-in-one medical image restoration via task-adaptive routing,'' in \emph{International Conference on Medical Image Computing and Computer-Assisted Intervention}.\hskip 1em plus 0.5em minus 0.4em\relax Springer, 2024, pp. 67--77.

\bibitem{deng2024unsupervised}
S.~Deng, Y.~Chen, W.~Huang, R.~Zhang, and Z.~Xiong, ``Unsupervised domain adaptation for em image denoising with invertible networks,'' \emph{IEEE Transactions on Medical Imaging}, 2024.

\bibitem{li2022annotation}
H.~Li, H.~Liu, Y.~Hu, H.~Fu, Y.~Zhao, H.~Miao, and J.~Liu, ``An annotation-free restoration network for cataractous fundus images,'' \emph{IEEE Transactions on Medical Imaging}, vol.~41, no.~7, pp. 1699--1710, 2022.

\bibitem{geng2021content}
M.~Geng, X.~Meng, J.~Yu, L.~Zhu, L.~Jin, Z.~Jiang, B.~Qiu, H.~Li, H.~Kong, J.~Yuan \emph{et~al.}, ``Content-noise complementary learning for medical image denoising,'' \emph{IEEE transactions on medical imaging}, vol.~41, no.~2, pp. 407--419, 2021.

\bibitem{wu2022arbitrary}
Q.~Wu, Y.~Li, Y.~Sun, Y.~Zhou, H.~Wei, J.~Yu, and Y.~Zhang, ``An arbitrary scale super-resolution approach for 3d mr images via implicit neural representation,'' \emph{IEEE Journal of Biomedical and Health Informatics}, vol.~27, no.~2, pp. 1004--1015, 2022.

\bibitem{dosovitskiy2020image}
A.~Dosovitskiy, L.~Beyer, A.~Kolesnikov, D.~Weissenborn, X.~Zhai, T.~Unterthiner, M.~Dehghani, M.~Minderer, G.~Heigold, S.~Gelly \emph{et~al.}, ``An image is worth 16x16 words: Transformers for image recognition at scale,'' \emph{arXiv preprint arXiv:2010.11929}, 2020.

\bibitem{zhang2021transct}
Z.~Zhang, L.~Yu, X.~Liang, W.~Zhao, and L.~Xing, ``Transct: dual-path transformer for low dose computed tomography,'' in \emph{Medical Image Computing and Computer Assisted Intervention--MICCAI 2021: 24th International Conference, Strasbourg, France, September 27--October 1, 2021, Proceedings, Part VI 24}.\hskip 1em plus 0.5em minus 0.4em\relax Springer, 2021, pp. 55--64.

\bibitem{zhang2024distilling}
Q.~Zhang, X.~Liu, W.~Li, H.~Chen, J.~Liu, J.~Hu, Z.~Xiong, C.~Yuan, and Y.~Wang, ``Distilling semantic priors from sam to efficient image restoration models,'' in \emph{Proceedings of the IEEE/CVF Conference on Computer Vision and Pattern Recognition}, 2024, pp. 25\,409--25\,419.

\bibitem{shen2019human}
Z.~Shen, W.~Wang, X.~Lu, J.~Shen, H.~Ling, T.~Xu, and L.~Shao, ``Human-aware motion deblurring,'' in \emph{Proceedings of the IEEE/CVF International Conference on Computer Vision}, 2019, pp. 5572--5581.

\bibitem{miao2024cross}
J.~Miao, C.~Chen, K.~Zhang, J.~Chuai, Q.~Li, and P.-A. Heng, ``Cross prompting consistency with segment anything model for semi-supervised medical image segmentation,'' \emph{arXiv preprint arXiv:2407.05416}, 2024.

\bibitem{wang2024sam}
C.~Wang, H.~Chen, X.~Zhou, M.~Wang, and Q.~Zhang, ``Sam-ie: Sam-based image enhancement for facilitating medical image diagnosis with segmentation foundation model,'' \emph{Expert Systems with Applications}, vol. 249, p. 123795, 2024.

\bibitem{zhang2023customized}
K.~Zhang and D.~Liu, ``Customized segment anything model for medical image segmentation,'' \emph{arXiv preprint arXiv:2304.13785}, 2023.

\bibitem{li2024carotid}
X.~Li, X.~Ouyang, J.~Zhang, Z.~Ding, Y.~Zhang, Z.~Xue, F.~Shi, and D.~Shen, ``Carotid vessel wall segmentation through domain aligner, topological learning, and segment anything model for sparse annotation in mr images,'' \emph{IEEE Transactions on Medical Imaging}, 2024.

\bibitem{gu2024lesam}
Y.~Gu, Q.~Wu, H.~Tang, X.~Mai, H.~Shu, B.~Li, and Y.~Chen, ``Lesam: Adapt segment anything model for medical lesion segmentation,'' \emph{IEEE Journal of Biomedical and Health Informatics}, 2024.

\bibitem{qiu2023large}
J.~Qiu, L.~Li, J.~Sun, J.~Peng, P.~Shi, R.~Zhang, Y.~Dong, K.~Lam, F.~P.-W. Lo, B.~Xiao \emph{et~al.}, ``Large ai models in health informatics: Applications, challenges, and the future,'' \emph{IEEE Journal of Biomedical and Health Informatics}, 2023.

\bibitem{jiang2024glanceseg}
H.~Jiang, M.~Gao, Z.~Liu, C.~Tang, X.~Zhang, S.~Jiang, W.~Yuan, and J.~Liu, ``Glanceseg: Real-time microaneurysm lesion segmentation with gaze-map-guided foundation model for early detection of diabetic retinopathy,'' \emph{IEEE Journal of Biomedical and Health Informatics}, 2024.

\bibitem{ma2024segment}
J.~Ma, Y.~He, F.~Li, L.~Han, C.~You, and B.~Wang, ``Segment anything in medical images,'' \emph{Nature Communications}, vol.~15, no.~1, p. 654, 2024.

\bibitem{li2024sam}
S.~Li, M.~Liu, Y.~Zhang, S.~Chen, H.~Li, Z.~Dou, and H.~Chen, ``Sam-deblur: Let segment anything boost image deblurring,'' in \emph{ICASSP 2024-2024 IEEE International Conference on Acoustics, Speech and Signal Processing (ICASSP)}.\hskip 1em plus 0.5em minus 0.4em\relax IEEE, 2024, pp. 2445--2449.

\bibitem{hinton2015distilling}
G.~Hinton, O.~Vinyals, and J.~Dean, ``Distilling the knowledge in a neural network,'' \emph{stat}, vol. 1050, p.~9, 2015.

\bibitem{murali2023endoscapes}
A.~Murali, D.~Alapatt, P.~Mascagni, A.~Vardazaryan, A.~Garcia, N.~Okamoto, G.~Costamagna, D.~Mutter, J.~Marescaux, B.~Dallemagne, and N.~Padoy, ``The endoscapes dataset for surgical scene segmentation, object detection, and critical view of safety assessment: Official splits and benchmark,'' \emph{arXiv preprint arXiv:2312.12429}, 2023.

\bibitem{lim2017enhanced}
B.~Lim, S.~Son, H.~Kim, S.~Nah, and K.~Mu~Lee, ``Enhanced deep residual networks for single image super-resolution,'' in \emph{Proceedings of the IEEE Conference on Computer Vision and Pattern Recognition Workshops}, 2017, pp. 136--144.

\bibitem{zamir2022restormer}
S.~W. Zamir, A.~Arora, S.~Khan, M.~Hayat, F.~S. Khan, and M.-H. Yang, ``Restormer: Efficient transformer for high-resolution image restoration,'' in \emph{Proceedings of the IEEE/CVF Conference on Computer Vision and Pattern Recognition}, 2022, pp. 5728--5739.

\bibitem{liang2021swinir}
J.~Liang, J.~Cao, G.~Sun, K.~Zhang, L.~Van~Gool, and R.~Timofte, ``Swinir: Image restoration using swin transformer,'' in \emph{Proceedings of the IEEE/CVF International Conference on Computer Vision}, 2021, pp. 1833--1844.

\bibitem{chen2022simple}
L.~Chen, X.~Chu, X.~Zhang, and J.~Sun, ``Simple baselines for image restoration,'' in \emph{European Conference on Computer Vision}.\hskip 1em plus 0.5em minus 0.4em\relax Springer, 2022, pp. 17--33.

\bibitem{jin2023let}
Z.~Jin, S.~Chen, Y.~Chen, Z.~Xu, and H.~Feng, ``Let segment anything help image dehaze,'' \emph{arXiv preprint arXiv:2306.15870}, 2023.

\end{thebibliography}
\end{document}